\documentclass[11pt,preprint]{aastex}

\usepackage{natbib}
\shorttitle{11 Mpc H$\alpha$ Survey}
\shortauthors{Kennicutt et al.}

\begin{document}


\title{An H$\alpha$ Imaging Survey of Galaxies in the Local 11 Mpc Volume}


\author{Robert C. Kennicutt, Jr.\altaffilmark{1,2}, 
Janice C. Lee\altaffilmark{2,3,4}, Jos\'e G. Funes, S.J.\altaffilmark{5,6}, 
Shoko Sakai\altaffilmark{7}, and Sanae Akiyama\altaffilmark{2}}


\altaffiltext{1}{Institute of Astronomy, University of Cambridge, 
Madingley Road, Cambridge CB3 0HA, UK}
\altaffiltext{2}{Steward Observatory, University of Arizona, Tucson, AZ 85721}
\altaffiltext{3}{Present address:  Carnegie Observatories, 813 Santa Barbara
Street, Pasadena, CA  91101}
\altaffiltext{4}{Hubble Fellow}
\altaffiltext{5}{Vatican Observatory, Steward Observatory, University of Arizona, Tucson, AZ 85721} 
\altaffiltext{6}{Visiting Astronomer, Cerro Tololo Inter-American Observatory.
CTIO is operated by AURA, Inc.\ under contract to the National Science
Foundation.}
\altaffiltext{7}{Division of Astronomy and Astrophysics, University 
of California, Los Angeles, Los Angeles, CA, 90095-1562}


\begin{abstract}

As part of a broader effort to characterize the population
of star-forming galaxies in the local universe, we have carried 
out an H$\alpha$+[NII] imaging survey for an essentially
volume-limited sample of galaxies within 11 Mpc of the Milky Way.  
This first paper describes the design of the survey, the observation,
data processing, and calibration procedures, and the characteristics
of the galaxy sample.  The main product of the paper is a catalog of
integrated H$\alpha$ fluxes, luminosities, and equivalent widths for 
the galaxies in the sample.  We briefly discuss the completeness
properties of the survey and compare the distribution of the sample
and its star formation properties to other large H$\alpha$
imaging surveys.  
These data form the foundation for a series of follow-up
studies of the star formation properties of the local volume, 
and the properties and duty cycles of star formation bursts
in dwarf galaxies.

\end{abstract}




\section{Introduction}

Much of our accumulated knowledge of the star formation
properties of nearby galaxies has been derived from the imaging, photometry,
and spectroscopy of galaxies in the H$\alpha$ nebular emission line
(e.g., Kennicutt 1998a and references therein).  CCD imaging observations
have proven to be especially valuable because they provide not only 
integrated measurements of star formation rates (SFRs), but also 
information on the spatial distribution and the luminosity
functions of individual star forming regions, with spatial resolutions
that are not easily matched at other wavelengths such as the ultraviolet
or far-infrared.  

The power of this technique is evidenced by the large number of 
published H$\alpha$ observations of galaxies, now comprising more
than 100 papers and nearly 3000 galaxies measured (Kennicutt et al. 2008, in 
preparation).  However this combined body of H$\alpha$ observations is 
heterogeneous, with strong biases favoring luminous, face-on, 
and high-surface brightness galaxies.  These provide representative
sampling of a diverse galaxy population, but do not 
constitute a statistically complete star formation inventory
of the local universe.


The most complete inventories
of local star-forming galaxies have come from spectroscopic surveys.
Objective prism surveys such as the Universidad Complutense de Madrid
(UCM) survey (Zamorano et al. 1994, 1996; Alonso et al. 1999), the 
Kitt Peak International Spectroscopic Survey (KISS;
Salzer et al. 2000, 2001; Gronwall et al. 2004, Jangren et al. 2005), 
and the CIDA--UCM--Yale survey (Bongiovanni et al. 2005)
provide H$\alpha$-based SFRs for redshift-limited samples over areas
of order $10^2 - 10^3$ deg$^2$ and survey volumes of order $10^5 - 10^6$ 
Mpc$^3$.  These offer emission line selected complete sampling of the most luminous 
and highest equivalent width (EW) H$\alpha$ emitters, and have yielded some 
of the most widely-applied measurements of the local comological SFR density
(e.g., Gallego et al. 1995).  Although such prism surveys have
well-defined sensitivity and detection properties across their survey volumes,
they become severely incomplete for galaxies with spatially extended low
surface brightness H$\alpha$ emission, and for galaxies with strong superimposed
continuua (i.e., low H$\alpha$ EWs).  As a result as much as half of
the aggregate star formation can be missed in these surveys 
(Sakai et al. 2004).  These problems can be reduced by undertaking
a comprehensive spectroscopic survey that uniformly targets galaxies within
a given magnitude and redshift range, such as the Sloan Digital 
Sky Survey (SDSS), and SDSS data have produced the most comprehensive
characterization of the low-redshift star forming galaxy population
to date (Brinchmann et al. 2004).  This approach has unmatched
statistical power, with spectroscopically measured H$\alpha$ fluxes
now available for hundreds of thousands of galaxies, but the data
still suffer from incompleteness introduced by aperture undersampling,
and the catalogs are heavily weighted to luminous spirals.

Another, complementary approach to this general problem is to 
obtain deep H$\alpha$ imaging for large and complete (or at least
robustly defined) samples of galaxies.  A number of ongoing surveys
are exploiting this approach.  These include two surveys of rich
galaxy clusters, the GOLDmine survey of the
Virgo cluster and Coma superclusters (Gavazzi et al. 2003), and
the MOSAIC H$\alpha$ survey, which has produced an H$\alpha$-selected
sample of star-forming galaxies in 8 $z < 0.03$ Abell clusters 
(Sakai et al. 2008, in preparation).  
Other surveys of (predominantly) field galaxies include
the H$\alpha$ Galaxy Survey (H$\alpha$GS; James et al. 2004), which
is based on a diameter and radial velocity selected sample of 334  
$z \le 0.01$ galaxies taken from the Uppsala Galaxy Catalog (Nilson 1973), 
the Survey for Ionization in Neutral-Gas Galaxies (SINGG; Meurer et al.
2006), which is based on an HI-selected subsample of 468 $z \le 0.0423$
galaxies from the HI Parkes All Sky Survey (HIPASS; Barnes et al. 2001),
the Analysis of the Interstellar Medium of Isolated Galaxies
(AMIGA; Verdes-Montenegro et al. 2005), which is observing a sample of
206 members selected from the Catalog of Isolated Galaxies 
(Karachentseva 1973), and a series of surveys of local 
irregular galaxies by Hunter and collaborators, culminating in
a survey of 140 galaxies by Hunter and Elmegreen (2004, hereafter
denoted as HE04).  

The most direct approach to assembling a truly complete
inventory of local star-forming galaxies would be to image {\it every}
galaxy within a fixed distance of the Milky Way.  Such a 
perfect census would
be free of selection biases and would provide a full characterization of 
the star-forming population, missing only those objects that are too
rare to appear in a small local volume.  A survey of this kind would provide
especially powerful constraints on the star formation and evolutionary
properties of low-mass dwarf galaxies, which will dominate the numbers
in any complete volume-limited survey.  Of course such an idealized survey
cannot yet be undertaken, because existing galaxy catalogs are incomplete
for dwarf galaxies even at small distances (e.g., 
Belokurov et al. 2007, Irwin et al. 2007, and references therein).
Within the Galactic zone
of avoidance this incompleteness increases and extends to brighter
luminosities.  Nevertheless one can achieve many of the scientific
advantages of a volume-complete survey by imaging the known 
galaxies in the local volume.  In this spirit we have undertaken to 
obtain and compile H$\alpha$ fluxes (and in most cases imaging) for 
most of the 261 known spiral and irregular galaxies 
within 11 Mpc of the Milky Way, above a Galactic
latitude $|b| = 20\arcdeg$, and with $B \le 15$, and for another
175 galaxies at lower latitudes and/or fainter magnitudes within this 
volume.  Within these limits the resulting sample
is virtually complete for M$_B \sim$ $-$15.5 and M(HI) $\sim 10^8$~M$_\odot$
(see \S5), and is statistically correctable for incompleteness to
luminosities and masses roughly an order of magnitude lower.
Another H$\alpha$ imaging survey of galaxies within
the local 10 Mpc volume is being carried out by Karachentsev and
collaborators (Kaisin \& Karachentsev 2007 and references therein),
and published data from that project have been used to
supplement our own observations.

Our resulting dataset provides a statistically rigorous sample for
characterizing the distributions of absolute and mass-normalized
star formation rates and 
HII region populations in the local volume.  The sample also allows us
to study the temporal variation of star formation, because in a 
volume-limited sample the relative numbers of galaxies observed in
various phases of a starbust cycle (for example), will scale with
the relative durations of those phases in the cycle.  
The sample also forms the foundation for two subsequent Legacy surveys
on the {\sl Galaxy Evolution Explorer (GALEX)} (Martin et al. 2005)
and the {\it Spitzer Space Telescope} (Werner et al. 2004).
The 11 Mpc Ultraviolet and H$\alpha$ Survey (11HUGS) is aimed at obtaining
deep ultraviolet imaging for most of the galaxies in the complete
($|b| = 20\arcdeg$, $B \le 15$) H$\alpha$ sample.  The 
{\it Spitzer}
Local Volume Legacy project (LVL) is aimed at obtaining 
infrared imaging at 3.6 -- 160 $\mu$m for the sample.  The combination
of H$\alpha$, ultraviolet, and infrared imaging of the galaxies
will provide extinction-corrected SFRs, and information on
the time dependence and spatial migration of the massive
star formation over timescales of $\sim10^7 - 10^8$ yr. 

This is the data paper for a series of papers that will use the H$\alpha$
observations of this sample to characterize the star formation 
properties of the local galaxy population (Lee et al. 2007), 
constrain the temporal behavior of star formation in dwarf galaxies 
(Lee 2006; Lee \& Kennicutt 2008, in preparation), 
and develop quantititative metrics of 
galactic star formation that can be applied to more distant galaxy samples.
In this paper we describe the design, construction, and execution of
the survey, and present the complete H$\alpha$ integrated flux and 
equivalent width (EW) catalog.  We also discuss the statistical
properties of the sample, in the context of other local narrowband
imaging surveys.

The remainder of the paper is organized as follows.  We first describe
the design and construction of the sample (\S 2).  In \S 3
we describe the observations, reduction, and calibration 
procedures.  The H$\alpha$ flux and emission-line equivalent
width catalog is presented in \S4, and we briefly describe the
properties of the galaxy sample and the dataset in \S5.  
We conclude the main paper in \S6 
with a brief description of the ancillary ultraviolet
and infrared observations now being obtained for the sample,
and the multi-wavelength public archive that will include the imaging
data presented in this paper.
Appendices provide details on the emission-line flux calibrations
and corrections for [NII] emission.

\section{Sample Definition}

The motivating scientific objectives of this survey were to 
characterize the full range of star formation properties among 
local galaxies, and to constrain the starburst duty cycle in 
low-mass systems.  Meeting these objectives dictated the design
of the survey sample.  The adopted distance limit of 11 Mpc
was large enough to ensure good sample statistics,
while avoiding the severe catalog incompleteness that is known to set in at
larger distances (e.g., Tully 1988b). 

As mentioned earlier, constructing a truly complete volume-limited survey of the
local volume is an unachievable ideal at the present time, because of 
incompleteness in current galaxy catalogs, and uncertainties in 
distances to the known galaxies, which make it impossible to define 
a precisely-bounded volume.  Instead we defined 
a distance-limited sample to the best of our current knowledge, and 
imposed additional limits on apparent magnitude and Galactic latitude 
that minimize the effects of incompleteness (and to enable corrections
for such effects when necessary).
It is well known that current inventories of nearby galaxies become 
severely incomplete below the Zwicky catalog limit (approximately
$B > 15$), and in the zone of avoidance.  Consequently we restricted
our primary sample to galaxies with $|b| \ge 20\arcdeg$ and 
$B \le 15$ (we address the resulting completeness properties of
this sample in \S5).  We also restricted the primary sample to spiral 
and irregular galaxies, i.e., those with Hubble types S0/a
and later, in order to avoid spending inordinate amounts of observing 
time imaging elliptical, dwarf spheroidal, or gas-poor S0 galaxies, 
which are known to contain few if any detectable HII regions (Pogge \&
Eskridge 1987, 1993, Kennicutt 1998a).  However we did include
early-type galaxies known to contain significant
levels of star formation (e.g., Cen A = NGC~5128).  

Thoroughout the remainder of this paper we designate as the ``primary sample"
of this survey the 261 galaxies that meet the combined criteria
of $d \le 11$ Mpc, $|b| \ge 20\arcdeg$, $B < 15$, and RC3 type $T \ge 0$.
However as observing time allowed we observed or compiled published 
data for other galaxies 
which fall outside of one or more of these limits, in what we
will designate as our "secondary sample."   These include 
galaxies below the primary sample magnitude and Galactic latitude limits,
and S0 galaxies that were observed to check whether significant levels
of star formation were present.
Since these data are useful for many other applications they are included in
this paper as well.  

Uncertainties in galaxy distances also complicate the task of
constructing a sample of this kind.  Thanks to several observational campaigns 
with the Hubble Space Telescope (HST) and groundbased telescopes,
direct stellar distances are being obtained for large numbers of
galaxies with distances within $\sim$5 Mpc (e.g., Karachentsev et al. 2004
and references therein; Tully et al. 2006), but they are available for only a handful
of objects near the survey distance boundary.  Thus one is forced
to estimate distances using either a less precise secondary indicator
(e.g., the Tully-Fisher relation) or by using the radial velocity 
of the galaxy and Hubble constant, corrected according to a
large-scale flow model.  Since our sample is dominated by dwarf
irregular galaxies, for which the Tully-Fisher relation is ill-defined,
we followed the lead of Kraan-Korteweg (1986) and Tully (1988a) and
used flow-corrected Hubble distances to select galaxies within 11 Mpc. 

Several flow models for the local volume are available, and we
devoted considerable effort to adopting one that is the best
compromise for estimating distances to galaxies within 11 Mpc.
For applications over larger volumes (e.g., $R \sim$ 15 -- 30 Mpc)
it is customary to use a single-attractor Virgocentric flow model
(e.g., Tully \& Shaya 1984, Mould et al. 2000).  However Karachentsev
\& Makarov (1996) have shown that for galaxies with $V_0 < 500$ km\, s$^{-1}$,
a pure Hubble flow with a dipole correction for Local Group motion
is consistent with direct distance measurements from Cepheids and other 
primary and secondary distance indicators.  However our survey extends
over larger volumes than those studied by Karachentsev \& Markarov (1996),
so the choice of models is not obvious.  In order to evaluate this ourselves
we compiled updated primary and secondary distances as available for
galaxies in our sample (see below), and compared them to distances
predicted by the Local Group and Mould et al. (2000) Virgocentric flow
models.  We found that the Karachentsev \& Makarov (1996) Local Group 
flow model provided more consistent distances over our 11 Mpc volume,
and hence we have adopted these flow distances when direct distance
measurements are not available.  Flow-corrected velocities for this
model are tabulated by the NASA/IPAC Extragalactic Database (NED),\footnote{The NASA/IPAC
Extragalactic Database (NED) is operated by the Jet Propulsion
Laboratory, California Institute of Technology, under contract
with the National Aeronautics and Space Administration.}
and we have applied them to estimate flow distances, using 
a Hubble constant of H$_{\circ}$= 75 km s$^{-1}$ Mpc$^{-1}$
(note that distances listed on NED use H$_{\circ}$= 73).



Candidate galaxies for our sample were drawn from the catalogs of
Kraan-Korteweg (1986), Tully (1988a), and Karachentsev et al. (2004; 
hereafter denoted KKHM), and augmented by galaxies in NED for which 
the flow-corrected velocities suggested a distance of 11 Mpc or less.
With this merged list of candidates in hand we then refined
the sample by accounting for available direct distance estimates
(from Cepheids, red giant branch tip, Tully-Fisher relation,
supernovae) for candidate galaxies and their groups, as compiled
from the literature (see references in Table 1).  Whenever possible
we adopted values already compiled in other papers, e.g., 
Ferrarese et al. (2000), Freedman et al. (2001), 
Tonry et al. (2001), Trentham \& Tully (2002), and KKHM.  
Group and cluster membership was generally taken from the Nearby Galaxies
Catalog of Tully (1988a; hereafter NBG), Bingelli et al. (1985),
Tully et al. (1996), and Trentham et al. (2001).  
The resulting set of galaxy and group distances allowed us to
exclude members of several groups and clusters as lying outside
of our sample.

{\bf Coma I group}:  The Coma I group (NBG cloud 14 group -1), which 
was previously assigned a distance of 9.7 Mpc in the NBG, is now known
to lie at an estimated distance of 16.4 Mpc (Trentham \& Tully 2002). 
Moreover one of its members, NGC 4414, has a Cepheid distance 
of 17.70 Mpc  (Freedman et al. 2001).  Thus, the 17 spiral and irregular
galaxies associated with this structure were removed from our 
candidate list. 

{\bf Ursa Major cluster}:  All members of the Ursa Major cluster
as identified by Tully et al. (1996) were excluded (18.6 Mpc;
Tully \& Pierce 2000), including NGC~3985, which was not included
as a cluster member in the NBG.

{\bf Virgo cluster}:  All Virgo cluster members were excluded
based on the compilations of Bingelli et al. (1985) and the NBG,
including NGC~4517, which was not listed as a member in the NBG,
but was subsequently identified as a cluster member by Gavazzi 
et al. (2002).

{\bf Fornax cluster}:  All Fornax cluster members were excluded
based on a mean Cepheid distance of 19.0 Mpc (Freedman et al. 2001).

{\bf NGC 3184 group}:  Four galaxies associated with the NGC 3184 group 
(NBG cloud 15 group +7) by Ferrarese et al.
(2000) were excluded based on Cepheid measurements of two of its members, 
NGC 3319 and NGC 3198, which yield a group distance of 
$\sim$13.6 Mpc (Freedman et al. 2001).

{\bf Leo cloud}:  NBG cloud 21 was excluded, based on a surface
brightness fluctuation (SBF) distance of 20 Mpc (Tonry et al. 2001).

{\bf Dorado (NGC 1566) group}:  NBG group 53-1 was placed outside
of the sample limits, based on a fundamental plane distance of
13.8 Mpc (Freedman et al. 2001).

{\bf Individual Cepheid distances}:  Four galaxies with flow-corrected
estimated distances of $<$11 Mpc were removed based on Cepheid distances:
NGC 2090 and NGC 2541 (11.75 Mpc and 11.22 Mpc, respectively, 
Freedman et al. 2001), NGC 2841 (14.1 Mpc, Macri et al. 2001), and
NGC 1637 (11.7 Mpc; Leonard et al. 2002).

On the other hand, direct distance estimates allowed us to 
firmly place two other groups within the 11 Mpc sample, even
when the flow velocities yielded distances near or above the limit:

{\bf NGC 1023 group}:  NBG group 17-1 was placed at a distance
of 9.2 Mpc, based on Cepheid measurements of NGC~925 (Freedman
et al. 2001).

{\bf Leo (NGC~3379) group}:  NBG group 15-1 was placed at a distance
of 10.0 Mpc, based on Cepheid measurements of NGC~3351 and NGC~3368
(Freedman et al. 2001).


The final step in the construction of the primary sample was
to compile $B$ magnitude measurements and impose a cut at 
$B = 15$.  The photometry sources are given in the references
to Table 1, with precedence given to the dwarf galaxy surveys
of van Zee, Haynes \& Salzer (e.g. van Zee et al. 1997)
and Binggeli, Barazza, Bremnes, Parodi, and 
Prugniel (e.g. Parodi et al. 2002), the RC3 (de Vaucouleurs et al 1991), 
and $B_T$ magnitudes compiled and reduced to the RC3 system by HyperLeda
(Paturel et al. 2003).\footnote{We acknowledge the usage of the 
HyperLeda database (http://leda.univ-lyon1.fr).}  
For 38 galaxies without such measurements, $B$ magnitudes were extracted
from other sources, or taken from NED.
The end result of this selection is a sample of 261 galaxies 
satisfying the primary sample criteria of $d \le 11$~Mpc, 
$|b| \ge 20\arcdeg$, $B \le 15$ and $T \ge 0$, and an additional 
175 galaxies that fall into the secondary sample.
 
We re-emphasize that our primary 11 Mpc sample as we have defined it only
approximates a clean, truly volume-limited $d \le 11$ Mpc data set.
More precisely, our sample should be regarded as consisting of objects with Local Group
flow-corrected radial velocities of $V \le 825$ km~s$^{-1}$,
with corrections applied for contaminants from more distant groups
and clusters.  To what extent does it approximate a true sampling
of an 11 Mpc volume?  The uncertainties in flow-corrected distances are
dominated by the dispersion in the local Hubble flow, which has
been measured for galaxies within 10 Mpc to be in the range 
60 $-$ 100 km~s$^{-1}$ (Sandage 1986, Karachentsev \& Makarov 1996, 
Whiting 2003, and references therein).  This implies distance
uncertainties of $\sim$7$-$12\% from random uncertainties alone.  
Systematic errors in the flow model are an additional source of uncertainty.
To estimate their effect, we compared the flow-corrected distances for our galaxies 
with those obtained using a Virgocentric infall model (Mould et al. 2000)
instead.  We found that the distances change by typically 1--2 Mpc,
with approximately 40 galaxies moving into or outside of the 
11 Mpc volume limit ($\sim$10\% of the entire sample).  These
are comparable to the random uncertainties arising from peculiar velocities,
and together they imply typical distance uncertainties of $\pm$15\%\ for
galaxies with flow-based distances.  Accordingly our 11 Mpc sample limit is
probably only hard at the level of $\pm$1.5--2 Mpc, and as distance
measurements and flow models are improved galaxies will migrate
in and out of the survey boundary.
Nevertheless, these caveats have minimal effect on the primary goal
of this survey, which is to characterize the star formation properties
of present-day galaxies using as close of a proxy as we can obtain
to a large, statistically useful volume-limited sample.

\subsection{Sample Characteristics}

Table 1 lists all 436 galaxies in the primary and secondary samples,
with identifications, positions, types, magnitudes, and distance
information.  This table is also available in machine-readable form
in the electronic edition.  Our follow-up {\it GALEX} and {\it Spitzer} surveys,
as well as the subsequent analysis papers in this series will be based on various
subsets of this master sample.  The table columns list the
following information:

Column (1):  The running index number in this table;

Column (2):  Galaxy name; 

Column (3):  Members of our primary and secondary samples are designated
             by p and s, respectively.  The primary sample is comprised
             of galaxies with $d \le 11$~Mpc,
             $|b| \ge 20\arcdeg$, $B \le 15$ and RC3 type $T \ge 0$.

Column (4):  J2000 right ascension, as reported in NED in 2005 June.  The
             nomenclature is hhmmss.s.

Column (5):  J2000 declination, as reported in NED in 2005 June.  
             The nomenclature is ddmmss.

Column (6):  Galactic latitude, as reported in NED in 2005 June.

Column (7):  Morphological type on the RC3 system (T-type), as reported
in the HyperLeda database in 2005 June.  
Classifications from NED were adopted for
12 dwarf galaxies that do not have assigned types in Hyperleda.
Further, blue compact dwarf galaxies (HII galaxies) often are misclassified
as elliptical galaxies.  Seventeen galaxies in our sample which also
are part of the Palomar/Las Campanas Altas of BCDs (Gil de Paz et al. 2003) 
were checked and reassigned a classification code of 11 (compact irregular).

Column (8):  Apparent $B$-band magnitude, with  
source listed in Column (14).

Column (9):  Corresponding uncertainty in $B$ magnitude;

Column (10):  Heliocentric recessional velocity, as reported in NED
in 2008 February.

Column (11):  Flow-field corrected recessional velocity, based on the
model described in \S 2. 

Column (12):  Adopted distance in Mpc.  Direct measurements from 
the literature were adopted when available, and were mostly taken from the
KKHM compilation, or from measurements
reported in Ferrarese et al. (2000) and Freedman et al. (2001).  
If direct measurements were not available, distances were computed 
from the flow corrected velocities listed in Column (11), using
H$_{\circ}$= 75 km s$^{-1}$ Mpc$^{-1}$.  The source of the distance
is referenced in Column (14).

Column (13):  The methods used to determine the tabulated distances,
with the following abbreviations:  Cepheid variables (ceph),
tip of the red giant branch (trgb), surface brightness fluctuations 
(sbf), membership in a group with a direct distance measurement (mem), 
brightest blue stars (bs), Tully-Fisher relation (tf), and
Virgocentric flow model distances (flow).

Column (14):  References for distances and magnitudes, as 
listed in the reference notes at the bottom of the table.  Distance
references are numbered 1--62 and photometry references are 
numbered 100--122.

\section{Observations and Image Processing}

\subsection{Observing Procedures}

Most of the H$\alpha$ and $R$ imaging reported in this paper was obtained
in 2001 $-$ 2004 using CCD direct imagers on the 
the Steward Observatory Bok 2.3~m telescope
on Kitt Peak (Bok), the Lennon 1.8~m Vatican Advanced Technology Telescope
(VATT), and the 0.9~m telescope at Cerro Tololo Interamerican Observatory
(CTIO).  Table 2 summarizes the properties of these instruments.
Common procedures were followed for all of the observations
except as noted below.

Data for most of the northern galaxies in the survey were
obtained with the 2K CCD imager at the Bok telescope, over 17
nights in 2001 March -- 2002 March.  The 
narrowband imaging was obtained using an 88~mm Andover 3-cavity interference
filter specifically designed for this project.  The combination of a high peak
transmission (90\%) and high (94\%) detector quantum efficiency
produced a high system throughput that
allowed us to achieve relatively deep flux and surface brightness limits
($\sim$$2\times10^{-16}$ ergs~cm$^{-2}$~s$^{-1}$ and 
$\sim4\times10^{-18}$ ergs~cm$^{-2}$~s$^{-1}$~arcsec$^{-2}$, respectively) 
in exposure times of only 1000~s.  In order to remove the stellar
continuum contribution to these images we also observed 
the same fields with a Kron-Cousins $R$ filter, with standard integration
times of 200~s, deep enough to provide supplemental broadband imaging
along with continuum subtraction for the narrowband images.  
Using the broadband continuum filter complicates the
data reduction, due to the significant mean bandpass shift with the
narrowband filter (Table 2), and the presence of H$\alpha$, [\ion{N}{2}],
and [\ion{S}{2}] emission lines in the $R$ band.  
These effects can be readily accounted for in the 
data reduction (Appendix A), and this produced continuum-subtracted images and
fluxes with the requisite accuracy for this survey.

Most of the observations on the VATT telescope were made using 
the same narrowband H$\alpha$ filter 
(a matching filter centered at 6600 \AA\ used for a few objects).
Longer total integration times of 1800~s (narrowband) and 360~s ($R$) 
were used to compensate for the smaller telescope aperture and the
somewhat lower quantum efficiency of the CCD detector, and yielded
the same signal/noise limits as the Bok observations to within 10\%.

Southern galaxies that were not accessible from Arizona 
were observed using the Cassegrain Focus
CCD Imager (CFCCD) on the CTIO 0.9~m telescope.  Data were obtained 
during 3 observing runs in 2001--2002.
A 75~\AA\ bandpass H$\alpha$ interference filter from CTIO was used
for the observations.  Because of the much smaller telescope aperture
it was not practical to achieve the depth of the Bok and VATT
observations, so exposure times were chosen (2700~s narrowband, 
300 -- 600~s broadband) to achieve approximately one third of the effective
exposure time.  The smaller aperture of this instrument was offset
by the wide field of view of the
CFCCD camera (13\farcm5), which allowed many of the largest galaxies in 
the project to be imaged efficiently.

Whenever possible the entire H$\alpha$-emitting disk of each galaxy
was imaged, if necessary using multiple pointings.  Calibration
exposures included zero, dome flatfield, twilight sky flatfield, and
dark exposures following standard practice.  Observations
of galaxies were interspersed with measurements of spectrophotometric
standard stars from the catalogs of Massey et al.
 (1988), Oke et al. (1990), and Hamuy et al. (1992, 1994) several times 
per night (under photometric conditions),
to calibrate the flux zeropoints. Most observations were made under photometric
conditions, but some data were taken through thin clouds, and were 
subsequently calibrated using short (300 -- 600~s) bootstrap exposures.
This process provided a measurement of the transparency at
the time of the first observations, and only observations taken with
a transparency of 50\%\ or higher were retained (in most cases the 
transparency was $>$80\%).    

\subsection{Image Processing and Continuum Subtraction}

Raw images were reduced following standard procedures using
IRAF\footnote{The Image Reduction and Analysis Facility (IRAF) is
distributed by the National Optical Astronomy Observatories, which are
operated by AURA, Inc.\ under cooperative agreement with the National
Science Foundation.}.  
Cosmic rays were excised using the JCRREJ2 
package (Rhoads 2000).

As with most CCD narrowband observations, the overall accuracy of our 
H$\alpha\ +$ [\ion{N}{2}] line maps are mainly limited by 
the fidelity of the flatfielding and (in weak line emitters) the
accuracy of the continuum removal (photometric zeropoint errors are
discussed separately below).  Standard flatfielding corrections using 
a combination of dome and twilight sky exposures produced images that
usually were flat to $\pm$1--3\%.  However 
all three imagers showed systematic structure in the sky background
arising from a combination of imperfect baffling, vignetting, and
scattered light in various combinations.  We were able to reduce some
of these residual instrumental features through use of dark sky flats.
These signatures have minimal effect on the integrated photometry
for most objects, but special care was taken to quantify their effects
on observations of galaxies with extended low surface brightness emission.

The processed narrowband images contain contributions both from 
H$\alpha$ and [\ion{N}{2}] line emission as well as underlying stellar
continuum (including H$\alpha$ absorption), and the accuracy of
the continuum scaling and subtraction can be the dominant error source
for galaxies with strong continuum and relatively weak H$\alpha$
emission (i.e., those with low emission-line EWs).  Net emission-line images
were obtained by subtracting a scaled $R$ image from the narrow-band image, 
aligning the respective images using forground stars.  The $R$-band scaling
factor for each instrumental setup was determined as follows.  
For galaxies with strong continua observed under photometric conditions,
continuum-subtracted images were computed for a range of scaling factors
and iterated; convergence was reached when the 
surface brightness of the continuum-dominated regions of the galaxy 
agreed with level of the background, and residuals fluxes of foreground
stars also reached a minimum.  This process was performed independently
for different galaxies with any given filter combination, and produced
an average best scaling factor.  This scaling was then applied to all
other galaxies observed with the same setup under photometric conditions,
although slight adjustments were applied occasionally when they produced
obvious improvements in the continuum-subtracted images.  This process
automatically corrects the emission-line images for underlying stellar
H$\alpha$ absorption, apart from minor spatial variations in the
absorption strength as a function of local dominant stellar population.
The latter typically are $<$2 \AA\ in absorption EW, which in turn
corresponds to $<$3\%\ of the total continuum signal in a 70 \AA\
bandpass filter.  For galaxies observed under non-photometric 
conditions, the scaling factor was adjusted manually to account
for changes in transparency between the $R$ and narrowband exposures.

\subsection{Astrometry}

An astrometric solution for each pair of narrowband and $R$-band images
was calculated using the MCSZERO and MSCCMATCH tasks in the IRAF MSCRED
package.  These routines performed cross-correlation
between the positions of stars in the $R$-band image and equatorial sky
position of matching stars in the USNO-A2 catalog.  The same
calibration is assumed for the corresponding narrow-band image, which
was previously aligned to the $R$-band image for the purposes of
continuum subtraction.  The solutions are described using the standard
World Coordinate System (WCS) keywords, and have rms deviations 
of $<$0\farcs5 for
all images.  This level of accuracy is important for the multi-wavelength
applications that will presented in future papers in this series.

\subsection{Absolute Flux Calibration}

With observations of this depth and photometric quality the
accuracy of the H$\alpha$ emission-line photometry is often limited
by the instrumental zeropoint calibration.  Reduction practice for
narrowband photometry varies
considerably among papers in the literature, so we have 
documented our calibration procedures in detail in Appendix A.
Here we briefly outline the steps that are involved.

Magnitudes of the observed spectrophotometric standards stars on the 
standard system were obtained by integrating
their spectral energy distributions over the filter response functions.  
Photometric zeropoints (unit responses) were then calculated by
comparing these values with the instrumental magnitudes measured 
through aperture photometry, and using a standard atmospheric
extinction coefficient of 0.08 mag airmass$^{-1}$.  Mean zeropoints
over a run were computed by averaging data from nights when 
$\sigma_{ZP} <$ 0.02 mag.  The conversion of these photometric zeropoints
to an absolute emission-line flux scale is described in Appendix A.
Galaxy images
originally taken during nights which showed ZP variations larger than 
$\pm$0.02 mag, or taken during non-photometric
nights when no standards were observed, were instead calibrated with 
short bootstrap calibration observations, or by using published
fluxes, as described in \S3.8.  

Since we are calibrating monochromatic emission-line fluxes with
observations of continuum sources (standard stars), a correction
for the filter transmission at the wavelengths of the emission
features is required.  This included explicit treatment of the
presence of three emission lines in the filter bandpass
(H$\alpha$, [\ion{N}{2}]$\lambda$6548,
and [\ion{N}{2}]$\lambda$6583), and the variation of the 
[\ion{N}{2}]/H$\alpha$ ratio between galaxies (Appendix B).

When a broadband filter such as 
$R$ is used as a proxy for the continuum band,
additional corrections are needed to compensate for the 
presence of emission lines in the continuum bandpass, and 
the shift in mean effective wavelength between
the narrowband and $R$ filters.  These 
corrections are typically small, of order a few percent,
but they can dominate the systematic errors in the
flux scales for high-quality CCD imaging.

\subsection{Aperture Photometry, Integrated H$\alpha$ Fluxes and Equivalent Widths}

The reduction steps outlined above produced a set of flux-calibrated
H$\alpha$ images for the galaxies, and some of the 
future applications of these data will use the full two-dimensional data.
However for many applications the main parameter of interest
is the integrated H$\alpha$ emission-line flux, or the ratio of the 
H$\alpha$ flux to the underlying continuum intensity, i.e., the 
integrated H$\alpha$ emission equivalent width (EW).  
We have measured fluxes and EWs for all of the galaxies observed,
and tabulate them in this paper, along with measurements of other galaxies
taken from the literature.  Here we describe the procedures used  
to measure the integrated fluxes and EWs.

Aperture photometry was performed with the aid of the GALPHOT 
package\footnote{GALPHOT is a collection of scripts
in the IRAF/STSDAS environment first developed by W. Freudling and 
J.J. Salzer. The current version has been further
enhanced by members of the Cornell Extragalactic Group and is 
maintained by M.P. Haynes.} for IRAF as follows.
First, the boundaries of the galaxies in the continuum and 
emission-line images were individually marked by eye, and 
foreground stars and background galaxies (or residuals of these contaminants
in the continuum-subtracted line images)
were automatically masked outside of this region, using the 
GALPHOT MCLEAN routine.  Masking of objects
or residuals within the galaxy boundaries was carefully done 
by hand to ensure that HII regions were not removed inadvertantly.

Three different ``aperture configurations" were used to measure count rates, 
each designed to minimize the contribution
of the sky.  The method applied most
frequently employs a curve-of-growth analysis with a set of ten 
concentric elliptical apertures which extend from 
the outer regions of the galaxies to the sky.  The flux at which 
the growth curve becomes level was adopted 
as the total integrated instrumental flux.  Typically, the variation 
of the enclosed flux among at least three 
consecutive apertures at this point was less than $\pm$2\%.  
This procedure is effective when the radial extent of the 
emission is substantially less than the field of view of the image, 
so that there is ample blank sky available for 
determining the background level.  A different procedure was used 
when a galaxy filled the field of view, or when the
sky background region was affected by vignetting.  
In such cases an average background level was estimated 
from regions in the image that were the least affected by galaxy 
emission or vignetting, and the count rate was
measured from the entire frame.  Finally, a third method was
employed for a few galaxies in which the nebular emission was confined 
to a few widely-separated faint \ion{H}{2}\ regions, 
with little apparent diffuse emission between them.  In such
cases large-aperture photometry of the galaxies produces unacceptably
large uncertainties ($>$20\%), and instead the fluxes of the detected
\ion{H}{2}\ regions were measured individually and added together
to estimate the integrated H$\alpha$ flux of the galaxy.

\subsection{Uncertainties in Fluxes and Equivalent Widths}

We have performed a number of internal and external comparisons
to estimate the accuracy of our integrated emission-line
fluxes and EWs.  We consider separately the 
primary error terms in the photometry: flatfielding accuracy,
the accuracy and uniformity of the continuum removal, and 
the accuracy of the photometric zeropoints.
An additional uncertainty in the H$\alpha$
maps (as separate from the H$\alpha\ +$ [NII] maps) is introduced by
variability in the local [NII]/H$\alpha$ ratio, and this
is discussed separately in Appendix B.  

Uncertainties in the flatfielding process were described in \S2.2.
The corresponding uncertainties in the integrated fluxes are 
estimated to be $\pm$1--4\%\ in most cases.  They are most important
in low surface brightness galaxies, but fortunately most such objects
were much smaller than the detector field of view, so we were able
to characterize any nonflatness and determine an accurate 
background-subtracted flux.  

Photometric zeropoint uncertainties for more 
than 80\% of our measurements are less than $\pm$2\%, as determined
from the consistency of the standard star photometry.  In cases 
where we calibrated a deep nonphotometric image with a short
photometric exposure the zeropoint accuracy was measured by 
comparing the flux ratios of stars in the field; again this
yielded typical accuracies of $\pm$2--3\%.  

The reliability of the continuum scaling and subtraction is the 
dominant source of uncertainty in most of our measurements.
Their accuracy was estimated in 
two ways.  First, we selected a dozen galaxies spanning a range of fluxes, 
EWs, and emission distributions, and produced continuum-subtracted
emission-line images using a range of narrowband/continuum scaling
factors, to identify the points at which the line images were
clearly oversubtracted or undersubtracted.  We found that in favorable
cases, such as bright spirals with large continuum-dominated regions,
the continuum scaling could be determined to within approximately $\pm$3\%.
On photometric nights this scaling factor should be virtually constant,
so observations of these galaxies were used to determine a mean
scaling factor for the run which was applied generally to the
photometric observations.  The normalization was less certain in galaxies
with a relatively weak continuum or those exhibiting
strong diffuse H$\alpha$ emission (including many of the dwarf
galaxies), which were observed during nonphotometric conditions.
In those cases the uncertainties in the continuum scaling ranged
from $\pm$5\%\ to $\pm$15--20\%\ in a handful of worst cases.  
As a check on these estimated uncertainties, we examined the 
distribution of the differences between the average scaling
factor determined for each observational set-up and the actual 
scaling factors used to subtract the photometric data.
The dispersion in the differences was 5\%, which is consistent 
with the results of the first exercise.  Thus, 
we have used the fractional difference between the actual and 
average scale factor, or a minimum uncertainty of
5\%, to calculate the corresponding fractional error in the 
emission-line flux due to uncertainties in the continuum 
level within the apertures used to measure the narrowband 
images.   This results in a median total flux error of 12\%, 
with lower ($\sim$10\%) uncertainties for higher EW ($>$20\,\AA) 
systems and larger uncertainties ($\sim$20\%) for lower EW
($<$20\,\AA) systems (see Figure 1).  

Uncertainties in the EW measurements are driven by 
the same factors.  The main differences are that 
photometric zeropoint errors are not 
relevant, but errors in continuum normalization are magnified,
because (for example) an underestimate of the continuum level
will both artificially elevate the net H$\alpha$ emission-line
flux and lower the continuum value to which that flux is normalized
when measuring the EW.  Nevertheless the fractional uncertainties
in the EWs tend to follow the corresponding uncertainties in the
fluxes to within a few percent in most cases.  The main exceptions 
a handful of galaxies with bright superimposed foreground stars,
which introduce errors of up to a few tenths of a magnitude
in continuum flux.  We
illustrate the distributions of the total uncertainties in 
our measurements by plotting the fractional errors in our 
fluxes and EWs as functions of the measurements themselves
in Figure 1.

In some cases we used published photometry to calibrate
the flux zeropoints of our images (when observations were
made in nonphotometric conditions), or to provide integrated
fluxes and EWs directly for galaxies that we did not observe ourselves.  
In those cases we relied in the first instance on the uncertainties
quoted in the primary literature sources.  However we were also 
able in many cases to check the consistency of the published
fluxes with the general literature, and revise these uncertainty
estimates if necessary, as described below.

\subsection{External Comparisons}

Many of the galaxies in this sample have independent 
H$\alpha +$ [\ion{N}{2}] and/or H$\alpha$\ flux measurements
from the literature, and 
these provide a valuable external check on our own photometry.
The best comparison data set for this purpose is the spectrophotometric
atlas of nearby galaxies by Moustakas \& Kennicutt (2006; hereafter
MK06).  This atlas is based on 8\,\AA\ resolution integrated spectra
over the wavelength range 3650 -- 6950 \AA, and were obtained by
drift-scanning the spectrograph slit over as much of the optical
extent of the galaxies as practical.  
Emission-line fluxes
were extracted from these spectra after fitting a stellar synthesis
model to the underlying continuum; this process provides precise 
measurements, even for relatively low emission-line EWs, and includes
explicit correction for underlying stellar H$\alpha$\ absorption.

There are 77 galaxies in common between our sample and MK06, with 61 
of these having independently calibrated fluxes from 
our observations.  The left panel of Figure 2
compares the respective H$\alpha$+[NII] fluxes, with the
logarithmic fluxes themselves compared directly in the top
subpanel, and the residuals plotted below.  Galactic
foreground extinction corrections have not been applied 
to either data set for this comparison.  The MK06 spectra sometimes
undersampled the full H$\alpha$ disks, so this needs to be
incorporated into our comparison.  The filled 
circles denote galaxies where the full extent of the 
H$\alpha$\ emission was thought to be sampled by the spectral
scan, while open circles show instances where the MK06
data undersampled the H$\alpha$ disks.  To provide a 
more accurate comparison in these cases we 
remeasured fluxes from our images with rectangular apertures 
matching the MK06 driftscan observations.  We 
have overplotted the resulting fluxes using
square symbols.

Considering that the fluxes from our survey and MK06 datasets 
have been measured using entirely different techniques, the 
consistency between of the respective fluxes is excellent.
For the galaxies which have been fully covered 
by the spectral drift scans, the mean offset in fluxes is only
0.007 dex (MK06 spectra 1--2\%\ brighter) with an rms 
dispersion about the mean of $\pm$0.086 dex.  The expected systematic 
difference for galaxies that suffer from spectral aperture 
undersampling is also seen, with the MK05 fluxes
being 15\%\ fainter on average.  However when identical 
apertures are used (black points) this mean offset is 
reduced to 0.012 dex (MK spectra fainter by $<3$\%).
The rms difference between measurements of individual
galaxies is $\pm$0.083 dex ($\pm$0.066 dex around the best
fit to the residuals).  These dispersions are consistent 
with average flux uncertainties of 10-15\%\ in both datasets.

We compare the EWs in right panel of Figure 2.  
Although the correspondence between the measurements is
still good, the MK06 EWs are higher on average, regardless of whether  
matched or mismatched apertures are compared.  The mean offsets
between our EWs and the MK06 values
for the fully covered and aperture limited galaxies are 
10\%\ and 20\%, respectively.  It may seem odd that galaxies
which are completely enclosed by the spectral scans should 
exhibit such an offset, but since the resultant spectral 
apertures are generally set by the extents of the 
nebular regions, they often do not contain all of the 
continuum emission, which is more extended in galaxies on 
average.  On the other hand, the apertures used to measure 
the continuum light in our imaging include the entire stellar 
disk (when not restricted by the FOV of the detector),
and are larger than the narrowband apertures on average.  
Thus, the spectral continuum flux densities are typically 
lower than those measured from the imaging, leading to 
comparatively higher spectral EWs.  However, these 
differences can only be partly explained by aperture 
differences.  When we use the spectral apertures to remeasure 
our scaled $R$-band images, the relative mean offset of the 
MK EWs decrease, but are still 7\% higher than our 
EWs.  It is possible that the method we use to scale the 
$R$-band images systematically overestimates the 
continuum flux density, but this would also result in 
systematically fainter fluxes, which are not seen.  
Another plausible explanation is that the sky background 
levels are overestimated in some of the spectroscopic data, where 
the spectrograph slit ($\sim$3\farcm3) is too short to adequately 
sample the sky.  The rms dispersion about the mean difference 
between our re-measured EWs and the MK EWs is $\sim$20\%, 
which is consistent with the dispersion in the seen the flux 
comparison.

As an additional check, we have compared our fluxes and EWs
to those compiled from the literature.  In a separate project
(Kennicutt et al. 2008, in preparation; hereafter KAL08),
we have searched the literature for integrated photometry of
galaxies in H$\alpha$\ (and H$\alpha$ + [NII]) fluxes, and
have reduced these to a uniform system including 
standardized corrections for Galactic
extinction and [NII] contamination.
We have also used multiple measurements of galaxies to 
check the reliability of each published dataset, adjust its
overall photometric zeropoint if necessary, and reject 
highly discrepant measurements or datasets.  
Fluxes from the KAL08 compilation are available for $\sim$60\% 
of the galaxies that we have observed,
and we compare our measurements to those data 
in the left panel of Figure 3 (excluding the MK06 measurements).  
The mean zeropoints between our data and the literature agree
to within 1\%.  However there is a 
large dispersion ($\pm$0.15 dex) about the mean.
If nothing else this illustrates the non-trivial challenge
in measuring accurate narrowband emission-line fluxes of galaxies.

Published EW measurements are also available from KAL08 
for $\sim$40\% of the galaxies that we have observed, and these
are compared to our measurements in the right panel of 
Figure 3.  Again, there is good overall 
correspondence between the measurements.  For the higher EW 
systems (EW $\ga$ 15\,\AA), there is a small mean offset of 
0.013 dex (our data 3\% lower) and a rms dispersion of 
0.164 dex ($\pm$46\%).  However, our measurements of the galaxies 
with the lowest EWs (EW $\la$ 15\AA) are systematically lower by 30\%.  
The EW values from the literature in this subset are primarily 
drawn from the H$\alpha$ Galaxy Survey of James et al. (2004). 
Those authors report in their paper that their EWs are
30\% larger on average than other measurements in literature,
and we have not applied any corrections to those values in
making this comparison.  Given the consistency of our EWs 
with MK06 and other published studies we tentatively attribute
the offset at low EWs in Figure 3 to systematic 
uncertainties in the James et al. (2004) measurements.

In summary, based on both the comparisons against the 
integrated spectral data of MK06 and the literature compilation
of KAL08, we conclude that the methods we have used to flux 
calibrate, continuum subtract and measure our narrowband
imaging result in measurements that are on an overall flux 
scale that is accurate to within $\pm$3\%.  Our
reported EWs are also generally consistent with other 
measurements in the literature, but we cannot entirely rule out a small
($\ll$10\%) offset.

\subsection{Additional Fluxes from the Literature}

Based on the comparisons presented above we have chosen to adopt
our own integrated fluxes for the 231 galaxies observed in photometric
conditions, and another 50 galaxies with imaging that were calibrated via
bootstrap observations performed under photometric conditions.  
Imaging for another 25 galaxies was calibrated using published
fluxes from the KAL08 data set described above.  Taken together these
provide calibrated imaging and fluxes for 306 members of the 
sample.

We have been able to expand the set of flux measurements further
by adding integrated measurements from other sources as compiled
in KAL08.  As described above, we independently assessed the reliability of the
individual measurements, usually by comparing multiple measurements
of a single galaxy or by comparing the set of data in a given paper
with other datasets.  After these checks were completed we were
able to add measurements for 90 galaxies that we were 
unable to image ourselves, as well as EW measurements for 46 of
those objects. 
Many of these are faint ($B > 15$) galaxies that fell 
outside of our primary selection criteria.  Since these additional data may be
useful for many applications we include them in the flux catalog
presented here.

\section{The H$\alpha$\ Flux and Equivalent Width Catalog}

The primary product of this paper is a set of integrated H$\alpha$ + [\ion{N}{2}]
flux and EW measurements for galaxies in the 11 Mpc volume.
This is presented in Table 3 (which is also available in the
electronic edition as a machine-readable table).
To summarize, we have obtained narrowband H$\alpha$+[NII] 
and $R$-band images for 304 of the 436 galaxies
identified in Table 1, where 288 of these have been independently 
calibrated through our own observations.  Integrated measurements 
from the literature are available for the majority of 
the remaining galaxies, and are incorporated into the catalog.  
Overall, 94\% of the galaxies in the sample have fluxes or strong
upper limits, and 89\% have fluxes and EWs.

Only 23 galaxies (6\% of the sample) lack measurements; some of 
these have not been observed due to bright foreground stars which make
accurate photometry impossible.  The physical properties of this
subset of galaxies are not biased in any particular way, so the 
absence of these measurements should not affect conclusions regarding 
the star formation properties of the greater sample.  

Within the full sample of 436 galaxies 22 were not detected in 
H$\alpha$, and for these 5$\sigma$ point source detection limits
are listed in Table 3.  Many of these galaxies are gas-poor
lenticular galaxies that fall outside of the primary sample, 
for which the absence of detectable star formation is unsurprising.
However several gas-rich dwarf galaxies also are undetected, and
the properties of these objects will be addressed in future papers.

We have chosen to tabulate the combined fluxes of the H$\alpha$
and satellite [\ion{N}{2}]$\lambda\lambda$6548,6583 forbidden
lines because most of the measurements were made with narrowband
filters that were sufficiently wide to contain most of the emission
from all three lines.  However we also provide a measured or estimated
[\ion{N}{2}]/H$\alpha$ flux ratio for each galaxy, to allow the
net H$\alpha$ emission-line fluxes and EWs to be readily calculated.
The entries in Table 3 are organized as follows.

Column (1):  The running index number, identical to Table 1.

Column (2):  Galaxy name, as in Table 1.

Column (3):  Integrated H$\alpha$ $+$ [\ion{N}{2}] emission-line flux and its 
1-$\sigma$ uncertainty.  Units of fluxes are $\log$ ergs~cm$^{-2}$~s$^{-1}$.


Column (4):  Integrated H$\alpha$ + [\ion{N}{2}] emission-line equivalent width (EW)
and its 1-$\sigma$ uncertainty, in units of \AA.


Column (5):  Source of flux \& EW measurements, coded as follows:

\begin{itemize}

\item  1.1.  Measurements and errors are based on new observations which 
were made under photometric conditions (228 galaxies).

\item  1.2.  Measurements and errors are based on new observations which
were obtained initially in nonphotometric conditions, but were subsequently
calibrated with short exposures obtained under photometric conditions
(50 galaxies).  The photometric precision of these fluxes should be
excellent but the signal/noise in the images may be slightly degraded
relative to the main set of observations.

\item  1.3.  Primary measurements are new observations made under non-photometric
conditions, but with flux zeropoints bootstrapped from measurements in the
literature.  For these galaxies the listed EWs were measured from the 
our images (26 galaxies).

\item  2.1.  Flux and EW measurements were taken from the literature, using 
the compilation of KAL08.  Uncertainties are derived from the dispersion of
multiple measurements when these are available, or from the uncertainty
cited in the original source for single measurements.  Otherwise a 10\%\
estimated uncertainty (0.04 dex) is applied  (57 galaxies).  

\item  2.2.  Flux measurements were taken from the literature compilation 
of KAL08 as above.  However published EW measurements are not available.  
In these cases we used published H$\alpha$ + [NII] line fluxes and
broadband photometry to estimate the continuum flux and emission-line
EW, following procedures detailed in KAL08.  
These estimated equivalent widths have an uncertainty
of $\pm$25\%, based on comparisons to galaxies with directly measured
EWs.  (25 galaxies)

\item  2.3.  Flux values were adopted from the compilation of KAL08.  EWs
are not available and cannot be estimated because of the lack
of broadband photometry (24 galaxies).  Many of these galaxies
are recently identified dwarfs from the surveys of 
Karachentsev et al. (KKHM and references therein).

\end{itemize}

Column (6):  Telescope used for imaging, as listed in Table 2.
If no telescope is listed the measurements were taken from the
literature, as compiled by KAL08.

Column (7):  Spatial coverage code for the imaging.  

\begin{itemize}

\item   1.  The boundaries of the star-forming regions are clearly seen 
and the nebular emission has been completely sampled (266 galaxies).  

\item   2.  The main star-forming disk is enclosed by the field of view,
but faint outlying HII regions or low surface
brightness features may fall outside of the field.  In most cases
the missing flux components should contribute less than 5\%\ of the
listed flux (26 galaxies).

\item   3.  The images do not fully cover the main star-forming disk, so the
fluxes listed are lower limits (16 galaxies).  

\end{itemize}

Column (8):  $B$-band Galactic foreground extinction.  The values listed are an 
average of the values based on the maps of Burstein \&
Heiles (1982) and Schlegel et al. (1998), when both are available.  
We adopted independently measured foreground extinctions from the literature 
for three galaxies that lie near the Galactic plane (IC 10, Maffei 2, Circinus), 
because the Schlegel et al. values become unreliable in these regions.
Sources were Richer et al. (2001) for IC10, Fingerhut et al. (2003, 2007)
for Maffei 1 and 2, respectively, and Freeman et al. (1977) for Circinus.

Column (9):  Absolute $B$ magnitude, based on the (apparent) $B$ magnitude
and distance given in Table 1.  These are corrected
for foreground Galactic extinction as listed in Column (8).

Column (10):  The adopted integrated 
[\ion{N}{2}]$\lambda\lambda$6548,6583/H$\alpha$ ratio (sum of
both components).

Column (11):  Source for the adopted 
[\ion{N}{2}]$\lambda\lambda$6548,6583/H$\alpha$ ratio, 
coded as follows:

\begin{itemize}

\item  1.  Integrated spectrophotometry of the galaxies 
from Moustakas \& Kennicutt (2006) 
or Jansen et al. (2000); 

\item  2.  Spectrophotometry of HII regions in the galaxies, 
as compiled from van Zee \& Haynes (2006) and references therein.
In cases where multiple HII regions were observed we averaged
the [\ion{N}{2}]/H$\alpha$ values.

\item  3.  If measurements were not available from one of these sources,
then the ratio was estimated using an empirical scaling relation
between [\ion{N}{2}]$\lambda$6584/H$\alpha$ and $M_B$, as described in 
Appendix B.

\end{itemize}

Column (12): Integrated H$\alpha$ luminosity, in units 
$\log$ ergs~s$^{-1}$.  It is derived using the H$\alpha$ + [\ion{N}{2}] flux listed
in Column (9), the distance listed in Table 1, and Galactic foreground
extinction, assuming $A_{H\alpha} = 0.6\;A_B$, and the [\ion{N}{2}]/H$\alpha$
ratio listed in Column (10).  No corrections for
extinction internal to the galaxies themselves has been applied here.
Note that for galaxies in this volume typical uncertainties in distances
are of order $\pm$10--20\%, which introduces corresponding uncertainties
in absolute luminosities of roughly $\pm$0.07--0.15 dex.

\section{Physical and Star Formation Properties of the Sample}

The main objective of this survey, as with the associated {\it GALEX} 11HUGS
and {\it Spitzer} LVL surveys, is to provide accurate statistics
and maps of massive star formation for a virtually complete volume-limited
sample of galaxies in the Galactic neighborhood.  As such 
our survey is highly complementary to the large body of other 
published and ongoing H$\alpha$ imaging surveys (KAL08 and
references therein).  As a guide to potential users of the dataset,
we briefly summarize the ensemble properties of our galaxy sample,
and compare them with other published H$\alpha$ imaging surveys
of nearby galaxies.

\subsection{Sample Properties and Completeness}

The main defining characteristics of our sample are summarized
in Figure 4, which shows the distributions of T-type, 
B magnitude, Galactic latitude and distance.
In each case the shaded histograms show the relevant distributions
for the entire target catalog (Table 1), while the solid black
lines show the corresponding distributions for our primary sample
(i.e. for $|b| \ge 20\degr$, $B \le 15$, T$\ge$0 and D$\le$ 11 Mpc).
The 22 galaxies which lack H$\alpha$ flux measurements are also 
indicated.  About half of these are in the
primary sample (shaded in black) and half are not (cross-hatched).
The samples are dominated by dwarf irregular galaxies (T = 9, 10),
as would be expected for distance-limited datasets.
The distribution of galaxies in apparent magnitude and Galactic 
latitude turn over sharply above $B \sim 15.5$ and below $|b| \sim 10\arcdeg$,
respectively.  These partly reflect the limits chosen for the primary
sample, but they also reflect a decline in the existing catalogs
of galaxies below these limits. 

Our imposition of a $B = 15$
magnitude limit produces a corresponding absolute magnitude limit
that brightens from $M_B \sim$ $-$10 at 1 Mpc distance to $M_B \sim$ $-$15
at the 11 Mpc sample limit.  This is illustrated in Figure 5, 
which plots the absolute magnitudes and distances for the primary
(solid points) and secondary samples (open circles).  Superimposed on
the figure are lines of constant apparent magnitude at $B =$ 15 and 15.5.
This falloff in completness below the $B = 15.5$ is readily apparent.
This is an inherent limitation of our current inventories of the
local galaxy population.  Extending our H$\alpha$ imaging to include
all known galaxies at yet 
fainter magnitudes would add significantly to the sample size, as illustrated
by the larger number of open circles below $B = 15$, but most of
the severe incompleteness for $B > 15.5$ (the Zwicky catalog limit)
would remain.

Finally, we illustrate the completeness of the primary sample
within its target range of $M_B < -15$.  In Figure 6 we compare
the distribution of blue luminosities of our primary sample
with luminosity functions derived for spiral and irregular 
galaxies from the Second Southern Sky Redshift Survey (SSRS2;
Marzke et al. 1998), and the Arecibo HI Strip Survey 
(Zwaan et al. 2001).  The differences 
at the faint end reflect the observational challenges and 
resultant uncertainties that are inherent in constraining the
number densities of faint galaxies (e.g., Trentham \& Tully 2002).
The two luminosity functions were chosen to mimic the selection
criteria of our sample (i.e., late-type galaxies detected
in HI), and to bracket the range of published faint-end slopes
to the luminosity function.
The distribution for our sample falls squarely between these down
to $M_B$ = $-$14 to $-$15, below which completeness falls dramatically,
consistent with what we infer from Figure 5.  A more thorough
analysis of the optical and HI completeness properties of
the sample shows that our sample is complete to the 95\% level
for $M_B < -15$ and M(HI) $\ge 3 \times 10^8$ M$_\odot$ (Lee 2006,
Lee \& Kennicutt 2008).

\subsection{Comparisons to Other H$\alpha$ Surveys}

This survey complements the large collection of H$\alpha$ imaging
surveys in the literature, which have been heavily weighted
in favor of luminous spirals with much higher SFRs than are
typical of our dwarf-rich sample.  This is illustrated 
in Figures 7 and 8, which compare the distributions of 
absolute magnitude, RC3 type, H$\alpha$ + [\ion{N}{2}] luminosity,
and H$\alpha$ + [\ion{N}{2}] EW for our full sample
to three other large narrowband imaging surveys of nearby field galaxies,
the UGC-selected H$\alpha$ Galaxy Survey by James et al. (2004), 
HI-selected SINGG survey (first release) by Meurer et al. (2006),
and the optically-selected HE04 irregular galaxy survey of Hunter
\& Elmegreen (2004).\footnote{We note that even larger surveys of nearby clusters
of galaxies have been completed recently (e.g., Gavazzi et al. 2003
and references therein), but since our sample
is strictly a field galaxy survey we restrict our present
comparison to other field samples.}

Figure 7 compares the (Galactic extinction-corrected)
luminosity and type distributions of the
four survey samples.  When compared to the H$\alpha$GS and SINGG
samples there is a clear difference in weighting in our sample toward
lower luminosity galaxies, and dwarf irregular galaxies (T-types
9 and 10) in particular, reflecting the high abundance of such
galaxies in the general galaxy population.  
The preponderance of low-luminosity irregular galaxies in the HE04 sample 
simply reflects the design of that survey.  Figure 7 shows that our survey
is not exploring any component of the galaxy population that
has not already been studied in earlier surveys; its primary
value rather lies in its homogeneous coverage of a local
sample with well defined limits in distance and magnitude.

Figure 8 compares the same surveys in terms of the distribution
of emission-line luminosities (which scale with the SFR) and
EWs (which scale roughly with the SFR per unit stellar mass).
The luminosities were corrected for Galactic extinction but not for
internal extinction in the galaxies themselves.
The trends seen in the H$\alpha$ + [\ion{N}{2}] luminosities
(left panels) mirror those seen in the absolute blue magnitudes.
Existing H$\alpha$ surveys cover virtually all of the range of
SFRs seen in our 11 Mpc sample, but the latter distribution
is skewed to lower luminosities relative to the H$\alpha$GS
and SINGG samples, as expected given the larger representation
of dwarf galaxies.   As expected there is good correspondence
between the distribution of luminosities in the HE04 sample
and the low-luminosity component of our sample;
this partly reflects a large overlap between the samples.
Galaxies within 11 Mpc comprise 57\%\ of the HE04 sample, compared
to 32\%\ and 35\%\ of the H$\alpha$GS and SINGG samples, respectively.

We can apply a very simple scaling of emission-line luminosities
to roughly characterize the range of SFRs in the sample.  For the H$\alpha$ 
SFR calibration of
Kennicutt (1998a) and a nominal [\ion{N}{2}]/H$\alpha$ ratio
of 0.2 (appropriate for the average luminosity of our 11 Mpc 
sample), an SFR of 1 M$_\odot$\,yr$^{-1}$ corresponds
to a logarithmic H$\alpha$ + [\ion{N}{2}] luminosity of
$\sim$41.2, for the units shown in Figure 8.  This means
that the galaxies in our sample have SFRs ranging from 
$\sim$10$^5$ -- 2 M$_\odot$\,yr$^{-1}$, without any correction
for internal dust attenuation.  Many of the more massive galaxies
have disk-averaged extinctions of up to 2--3 mag at H$\alpha$
(e.g., Calzetti et al. 2005, Moustakas \& Kennicutt 2006),
so the range in attenuation-corrected SFRs is closer to 
$\sim$10$^5$ -- 10 M$_\odot$\,yr$^{-1}$.  

An important result that may not be  
immediately apparent from Figure 8 is the extreme rarity
of spiral and irregular galaxies which are completely devoid
of H$\alpha$ emission.  The entire sample contains 22 H$\alpha$
non-detections, with many of those being   
elliptical, SO, or dwarf spheroidal galaxies that
were observed simply to confirm the absence of line emission.
The remainder save one are all extreme dwarf irregular galaxies with
$M_B >$ $-$13 and M(HI) $< 5 \times 10^7$ M$_\odot$ (the
exception, UGC 7356, has $M_B = -13.6$ and M(HI) $< 5 \times 10^8$ M$_\odot$).  
The upper limits on the H$\alpha$ luminosities
in these objects are of order a few times $10^{35}$ ergs\,s$^{-1}$,
corresponding to instantaneous SFRs 
(if the standard conversion factors were applicable
in this regime) of order a few times $10^{-6}$ M$_\odot$\,yr$^{-1}$.
However these SFRs are not physically meaningful, because the 
assumptions that underlie the use of a global H$\alpha$ to SFR
conversion, including a fully populated stellar initial mass 
function and a steady SFR over the past 5--10 Myr, do not
apply in this regime (e.g., Cervi\~no et al. 2003;
Weidner \& Kroupa 2006).  Nevertheless the results 
underscore the sensitivity of our survey (and the 
corresponding survey limits).  Our result is consistent with an earlier
finding in the SINGG survey, which detected H$\alpha$ emission
in all 93 of its first-release sample (Meurer et al. 2006).
As discussed in that paper, massive star formation appears
to be nearly ubiquitous in present-day spiral and irregular
galaxies, at least down to systems with HI masses of
$10^7 - 10^8$ M$_\odot$.  This topic is addressed in more
depth in future papers in this series.

The righthand panels in Figure 8 compare the distributions
of H$\alpha$ + [\ion{N}{2}] emission-line EWs for the four
samples.  Again we removed any corrections for internal
extinction in the published data.  The HE04 paper does not
include EW measurements, so we estimated the EWs using
a combination of the published H$\alpha$ fluxes and broadband
photometry of the galaxies, as described in KAL08.  Individual
synthetic EWs are only accurate to $\pm$20\%, but are sufficient
for comparing the general distribution of EWs to the other galaxy
samples.  Since the EWs are basically the ratio of the emission
line fluxes to the underlying red continuum fluxes, they 
roughly scale with the specific SFR, the SFR per unit stellar
mass.  A handful of galaxies in each sample have EWs of
$>$150\,\AA\ (exceeding 500\,\AA\ in extreme cases), and these
are grouped together in the last plotted bin to preserve
a reasonable range in the plots.  
The EW distributions for the four samples are surprisingly
similar, when considering the large differences in the
properties of the parent galaxy samples and their absolute
SFRs.  This is largely a manifestation of a very weak dependence
of average EW on galaxy luminosity or mass (Lee et al. 2007),
so that even large changes in characteristic mass do not
shift the EWs distributions significantly.  Our  
sample does show a considerably larger dispersion in the
distribution of EWs relative to the other three samples.
This reflects the diversity of its parent galaxy
population, which is much richer in dwarfs than the H$\alpha$GS
and SINGG samples and much richer in luminous spirals, which also
show a wide range in EWs, relative to the HE04 sample.
By contrast the H$\alpha$GS and SINGG samples are heavily
represented (relative to a volume-complete sample)
in intermediate-luminosity, late-type spiral galaxies,
which show the tightest distribution in EWs (Lee et al. 2007). 

The cosmic volume probed by our survey is relatively
small ($\sim$4000 Mpc$^3$ in the primary sample), so massive
galaxies are poorly represented.  As a result 
we would not necessarily expect the distribution of SFRs to
be representative of the local universe as a whole, at least
for the highest SFRs.  Nevertheless we can qualitatively compare the 
emission-line luminosity distribution to those derived from
large-volume objective prism emission-line surveys.
This comparison is presented
in Figure 9.  The top panel shows the distribution of 
H$\alpha$ luminosities for our primary sample,
taken as a whole (black lines), and including only those
galaxies with EW(H$\alpha$) $>$ 20 \AA\ (red lines).
The second set of lines is intended to approximately
replicate the insensitivity of the prism surveys to
galaxies with low emission-line EWs 
(Gronwall et al. 2004).  For each of these cases the dotted
lines show the observed H$\alpha$ luminosities without
a correction for extinction, while the solid lines
show luminosities with an approximate extinction correction
scaling with parent galaxy luminosity, following
the algorithm of Lee (2006):

\begin{equation}
\mbox{A(H$\alpha$)} = 0.14  ~~~~~~~~~~~~~~~~(M_B \ge -15)
\end{equation}

\begin{equation}
\mbox{A(H$\alpha$)} =  1.971 + 0.323 M_B + 0.0134 M_B^2  ~~~~~~(M_B < -15)
\end{equation}

\noindent
This is a crude statistical correction at best, but 
is far superior to adopting a single constant correction,
especially for a sample with such a wide range in parent
galaxy luminosities.

The bottom panel compares the H$\alpha$ luminosity distribution
of our sample (restricted
to galaxies with EW(H$\alpha$) $>$ 20\,\AA) to the corresponding 
luminosity functions published from the UCM (Gallego et al. 1995)
and KISS surveys (J.\,Salzer, private communication).
As before observed luminosities are plotted with dotted lines
and extinction-corrected luminosities are plotted with solid lines. 
The luminosity distribution for our sample falls between the two prism surveys
where the completeness limits overlap, though our  
data extend to fainter limits.  The main difference is the  
falloff in the luminosity function for the very brightest
galaxies in our sample (for $\log$L(H$\alpha > 42$ or SFR $>$ 5--10 
M$_\odot$\,yr$^{-1}$), where statistics are sparse due to the small
survey volume.  It is interesting that our H$\alpha$
luminosity function also flattens 
below L(H$\alpha$) $\sim 10^{40}$ ergs\,s$^{-1}$, or $\sim$0.1 M$_\odot$\,yr$^{-1}$),
whereas the corresponding luminosity functions from the
prism surveys are still rising at their completeness limits.
Some flattening of the luminosity function could result from
incompleteness in the local galaxy inventories; however we
would not expect such incompleteness effects to be important
at these (relatively high) SFRs.  Instead we suspect that 
this relatively shallow faint-end H$\alpha$ luminosity function 
is real, and that the steeper slopes in the prism surveys 
result from either small-volume statistics, photometric uncertainties,
and/or confusion between star-forming galaxies and individual
emission regions within galaxies at the faintest luminosities.

\section{Future Papers and Data Archive}

The main goal of this paper has been to present a homogenous
database of high-quality H$\alpha$ photometry and imaging
for a distance limited sample of galaxies in the local volume.
Future papers in this series will apply the data 
to characterize the distribution of star formation rates
in the local universe and to place quantitative constraints on the duty cycle of 
starbursts in dwarf galaxies.
We are also using these data in conjunction with other surveys
to study the Schmidt star formation law (Kennicutt 1998b) 
and other SFR scaling relations, and to calibrate metrics
of star formation activity for eventual application to lookback studies
(Kennicutt et al. 2005a, b).

As mentioned in \S1 the H$\alpha$ survey presented here is
being expanded (via larger collaborations) to the ultraviolet
({\it GALEX} 11HUGS survey), and the infrared ({\it Spitzer} LVL survey).
The combined multi-wavelength data set will be
applied to characterize the star formation, interstellar
dust, stellar populations, and evolutionary properties
of this unique sample of galaxies.  Applications
that will be enabled by these data include accurate measurements
of extinction, the measurement of more accurate SFRs at low
levels where the integrated H$\alpha$ emission becomes
unreliable, and the ability to constrain the temporal
behavior of the star formation over the past 0 $-$ 200 Myr.
A primary scientific product of these Legacy surveys will
be a multi-wavelength image atlas, that will include
the H$\alpha$ and $R$-band CCD images obtained for this study.
We currently anticipate delivery of the final data products
to the {\it Spitzer} Science Center in mid-2009.

\acknowledgements

We express our special thanks to Liese van Zee, Prasanth Nair, and
Kate Barnes for allowing us to include unpublished H$\alpha$ photometry 
of the newly discovered Leo T galaxy in this paper.  We are also  
grateful to a number of individuals for valuable discussions
and other contributions to the results presented here, including in particular
Ayesha Begum, Matt Bothwell, Caina Hao, Deidre Hunter, Gerhardt Meurer, Jeremy Mould, 
John Moustakas, Chien Peng, Evan Skillman, Christy Tremonti, and Liese van Zee.
We also are grateful to Matt Bothwell for assisting with the production
of the on-line tables for the paper.  Finally we are grateful to the anonymous
referee for their thoughtful reading and comments on the paper.
This work was supported in part by NSF grants AST-9900789 and AST-0307386,
and by NASA grants NAG5-8426 and GALEXGI-04-0048-0047.

\appendix
\section{Emission-line Flux Calibration}

Here we describe in detail the process used to convert observed
narrowband count rates to absolute emission-line fluxes.  These
apply to observations made with filter bandpasses of 65 $-$ 75 \AA,
which typically include contributions from H$\alpha$ as well as the
\ion{N}{2}\,$\lambda\lambda$6548,6583 doublet.  Our
reduction takes into explicit account the variation of the 
[NII]/H$\alpha$\ ratio between galaxies (described separately
in Appendix B), and the changes in filter transmission for 
each of the three emission lines with changing redshift.
We have followed the common practice of 
using a broadband $R$ filter for the continuum measurement, and
our reduction incorporates corrections for emission-line contamination
of the $R$ image and shifts between the mean wavelengths of the
the H$\alpha$ and $R$ filters.  Although the method is optimized
for the specific observing setup we use, we have generalized our
so the procedure can be applied more generally to narrow-band
emission-line imaging observations.

\subsection{Unit Response}

To derive a physical flux, the first step is to calculate the unit
response of the detector-filter-telescope combination, i.e., 
the ratio of the observed count rate, $CR$ [counts$\;\;$s$^{-1}$], 
to the mean source 
flux density, $f_{\lambda}$ [erg s$^{-1}$ cm$^{-2}$ \AA$^{-1}$].
For a given observation, the instrumental magnitude $m_{inst}$ is
related to the calibrated magnitude $m$ through the basic equation
\begin{equation}
m = m_{inst} - \kappa \sec(z) + ZP\;,
\end{equation}
\begin{equation}\label{m_inst}
 = -2.5 \log \mbox{CR} - \kappa \sec(z) + ZP\;.
\end{equation}

\noindent The photometric zero point $ZP$ is based on observations of 
spectrophotometric standards (\S3.1).  For our
reductions we assumed a constant atmospheric extinction
coefficient $\kappa$ of 0.08 mag airmass$^{-1}$ (we 
took pains to obtain standard star observations at similar
airmass to the target galaxies, so our calibrations are insensitive
to the adopted extinction coefficient).

Using the calibration from Massey et al. (1988),  
the monochromatic magnitude $m_{\nu}$ is given by

\begin{equation}
m_{\nu} = -2.5 \log f_{\nu} - 48.59\;.
\end{equation}

\noindent To express this in terms of $f_{\lambda}$, we apply the relation
$f_{\nu} = \frac{\lambda^2}{c} f_{\lambda}$, so that

\begin{equation}\label{m_nu}
m_{\nu} = -2.5 \log (\lambda^2 f_{\lambda}) - 2.397\;.
\end{equation}

\noindent Combining Eqs. \ref{m_inst} and \ref{m_nu} yields the unit response for the image: 
\begin{equation}\label{U}
U\;\;\Bigg[\frac{\mbox{counts s}^{-1}}{\mbox{erg s}^{-1}\;\mbox{cm}^{-2}\;\mbox{\AA}^{-1}}\Bigg] \equiv \frac{f_{\lambda}}{\mbox{CR}} = \lambda^{-2} 10^{-0.4(ZP +2.397-\kappa \sec(z))}\;.
\end{equation}
 
\hspace{0.05in}
\subsection{Transmission Corrections}
The next step is to compute T$(\lambda)$, the
effective transmission of the filter at the wavelength of the
redshifted H$\alpha$ and [\ion{N}{2}] lines.  
The total emission line flux, $f_{tot}$, 
as observed at the top of the Earth's atmosphere is

\begin{equation} \label{T}
f_{tot}(\mbox{H$\alpha$+[NII]}) = \frac{f_{cal}(\mbox{H$\alpha$+[NII]})}{\mbox{T}(\lambda)}\;, 
\end{equation}

\noindent so that $f_{cal}$ represents the calibrated line flux
as measured from the continuum subtracted narrowband image,
before the application of a transmission correction.

To produce images where only H$\alpha$+[NII] emission is present, the
$R$-band image is scaled and subtracted from the narrow-band image as
described in \S3.2.  However, emission-line flux is also contained within
the bandpass of the $R$ filter, so in this process of continuum
removal, a small fraction of the true H$\alpha$+[NII] flux is lost from
the narrow-band image.  We can recover this flux, and compute
T$(\lambda)$ along the way, by expressing the image subtraction
procedure algebraically:

\begin{equation}
 f_{tot}(\mbox{H$\alpha$+[NII]})\;\;T_{NB}(\lambda)\;\;t_{NB} - f_{tot}(\mbox{H$\alpha$+[NII]})\;\;T_{R}(\lambda)\;\;t_R\;\;\digamma^{-1}  = f_{cal}(\mbox{H$\alpha$+[NII]})\;\;t_{NB}\;,
\end{equation}

\noindent where $T_{R}(\lambda)$ and $T_{NB}(\lambda)$ are the
normalized filter transmission, as described at the end of this
section, $t_R$ and $t_{NB}$ are the exposure times, and $\digamma$ is
the ratio of counts for a continuum source between the $R$-band
and narrow-band observations (also the scaling factor applied
to the $R$-band image in producing a continuum-subtracted narrow-band
image).

Solving for $f_{tot}$(H$\alpha$+[NII]) results in an expression that
incorporates corrections for both the presence of the H$\alpha$+[NII]
in the $R$-band filter, and the differential transmission across both
$R$ and narrow-band filters:

\begin{equation}
f_{tot}(\mbox{H$\alpha$+[NII]}) = f_{cal}(\mbox{H$\alpha$+[NII]})\;\;\Bigg[T_{NB}(\lambda) - T_{R}(\lambda) \frac{t_R}{t_{NB}} \frac{1}{\digamma}\Bigg]^{-1}\;.
\end{equation}

\noindent Comparing this with Eq. \ref{T} we see that

\begin{equation}
T_{\lambda} = T_{NB}(\lambda) - T_{R}(\lambda) \frac{t_R}{t_{NB}} \frac{1}{\digamma}\;.
\end{equation}

\noindent The second term in this equation describes the
line flux that is lost during the subtraction of the continuum, and
amounts to an effective reduction in the line transmission by 4\%
for most of our galaxies.  A graphical illustration of this
correction is shown in the bottom two panels of Figure 10.
The curve which describes the
final transmission correction, $T_{\lambda}$ (solid line), is depressed
relative to $T_{NB}(\lambda)$ (open circles).  The size of the
effect is comparable to the ratio of the integrated throughputs
between the narrow and $R$-band filters (or in this case to
approximately the ratio of the bandwidths of the $R$ and narrow-band filters).

Finally, we need to determine $T_{R}$ and
$T_{NB}$, which are {\it not} simply the normalized
transmissions at the wavelength of the redshifted H$\alpha$ line.
Since the narrowband filters used are $\sim$70 \AA\ wide, emission from
[NII]$\lambda\lambda$6548,6583 is also present in the images, and these
flanking lines generally will not be attenuated by the same factor as 
the H$\alpha$
flux.  To account for this, the transmissions at the redshifted
wavelengths for all three lines are computed, and the average of these,
weighted by the relative fluxes of the lines, is taken.  The [NII]/H$\alpha$
ratios that we have used are given in Table 3, and a method for estimating 
them when spectral measurements are not available is described in Appendix B.

The filter transmission curves used for these calculations were
either those provided by the manufacturer (Bok, VATT observations),
or measured by NOAO (CTIO). For the multi-cavity interference filters used
in this project the characteristic filter shape
is an extended peak with roughly constant tranmission, extending
over 60--80\%\ of the FWHM bandpass, with steeply falling transmission
on either side.  This assured that the H$\alpha$ line fell at peak
transmission, in a region of roughly constant transmission with
wavelength, for all of our galaxies.  This also eliminated the
need for corrections for bandpass shifts introduced by the converging
beam or by temperature changes; that latter only amounted at most
to 1--2\%\ of the filter bandwidths in any case.

We compare the effect of using a
flux-weighted transmission correction 
with one that is simply read off from the normalized filter tracing at
the position of H$\alpha$ in Figure 10.
Naturally, the disparity between the two transmissions will 
increase as the target
galaxy's [NII]/H$\alpha$ line ratio increases.  We illustrate
the largest disparity possible, with [NII]$\lambda$6584/H$\alpha$=0.54.
The wavelength range over which H$\alpha$ is shifted for the
recessional velocities spanned by the galaxies in our sample is
marked by the the solid red and blue lines.  The corresponding
positions of the [NII]$\lambda\lambda$6548,6583 lines are also drawn in
the middle pannel.  In the $R$-band (upper panel) the difference
between these two transmissions are negligible ($0.2\%$) since the
throughput is essentially constant over the relevant wavelength range.
However, there are differences of a few percent in the narrowband
(middle panel) that vary according to the redshift of the target.  The
residual from the ``native'' transmission of the narrowband filter at the
shifted wavelength of H$\alpha$ is shown in the bottom panel.  


\subsection{The Final Calibrated Emission Line Flux and Equivalent Width}

We now combine the results from the last two sections to arrive at a 
final expression for $f_{tot}$.  The calibrated flux $f_{cal}$ 
(prior to corrections for differential transmission) can be 
computed by summing the unit response 
$U$ over the filter bandpass and then multiplying by the measured 
emission-line count rate from the continuum-subtracted narrow-band image:

\begin{equation}\label{f_cal}
f_{cal}(\mbox{H$\alpha$+[NII]}) =  U \cdot \mbox{FWHM$_{NB}$} \cdot 
\mbox{CR}(\mbox{H$\alpha$+[NII]}) \;.  
\end{equation}

Finally, the transmission correction $T_{\lambda}$ is applied through the 
combination of Equations \ref{T} and \ref{f_cal}, and $U$ is written out 
explicitly using Equation \ref{U}.  This yields:

\begin{equation}
f_{tot}(\mbox{H$\alpha$+[NII]}) =  U \cdot \mbox{FWHM$_{NB}$} \cdot 
\mbox{CR}(\mbox{H$\alpha$+[NII]})\;\;\Bigg[T_{NB}(\lambda) - T_{R}(\lambda) 
\frac{t_R}{t_{NB}} \frac{1}{\digamma}\Bigg]^{-1}\;,
\end{equation}

\begin{equation}
= \lambda^{-2}\;10^{-0.4(ZP +2.397-\kappa \sec(z))}\;\mbox{FWHM$_{NB}$}\;\; \mbox{CR}(\mbox{H$\alpha$+[NII]}) \;\Bigg[T_{NB}(\lambda) - T_{R}(\lambda) \frac{t_R}{t_{NB}} \frac{1}{\digamma}\Bigg]^{-1}
\end{equation}

The definition of the emission-line equivalent width (EW) is
straightforward:

\begin{equation}
EW(H\alpha+[NII]) = {{f_{tot}(H\alpha+[NII])} \over {f_{\lambda}(6563\AA)}}
\end{equation}


\begin{equation}
   = \frac{ \mbox{FWHM$_{NB}$} \cdot \mbox{CR}(\mbox{H$\alpha$+[NII]})
   \cdot
   (T_{\lambda})^{-1} \cdot t_{NB} }{\mbox{CR}(\mbox{Rnorm}) \cdot t_{R}
   - \mbox{CR}(\mbox{H$\alpha$+[NII]}) \cdot (T_{\lambda} \digamma)^{-1}
   \cdot t_{R}  \cdot T_{R}(\lambda)}
   \end{equation}

\noindent
where $f_{\lambda}$(6563\,\AA) is the continuum luminosity at the
rest wavelength of H$\alpha$.  In practice this was derived from the
uncalibrated emission-line and continuum images, taking advantage of
the fact that the normalized $R$-band image used for continuum removal
is scaled to the same photometric scale as the emission-line
image.  In that case the EW is determined solely by the ratio of
countrates in the emission-line and continuum image, multiplied
by the effective full-power width of the narrowband interference filter.


\section{[NII]/H$\alpha$ Estimation}

As discussed in \S3 spectrophotometric measurements of HII regions
in many of the galaxies in this sample, or even integrated spectra of the
galaxies themselves are available in the literature, and these
provided an accurate measurement of [\ion{N}{2}]/H$\alpha$ for
those objects.  For those galaxies without direct spectral
measurements of the [NII]/H-alpha ratio,
we apply a statistical correction based on
the B-band luminosity of the galaxy.

Figure 11 shows the scaling betwen [NII]$\lambda$6584/H$\alpha$ and
$M_B$, for a large set of integrated spectra of galaxies from 
Moustakas \& Kennicutt (2006; MK06).
The $M_B$ values
have been corrected for Galactic extinction but not internal extinction,
and were derived assuming $H_0=$75 km s$^{-1}$ Mpc$^{-1}$.  Filled
points represent galaxies whose line emission is due to star formation,
while open triangles are those in which the ionization is non-thermal,
probably from the presence of a bright active galactic nucleus (AGN).
Open squares show galaxies with ambiguous classifications.
All of these determinations were taken from MK06 and are based on the standard
[OIII]$\lambda$5007/H$\beta$ vs [NII]$\lambda6583$/H$\alpha$
diagnostic diagram (BPT; Baldwin, Phillips \& Terlevich 1981).  

A strong correlation between [\ion{N}{2}]/H$\alpha$ and $M_B$
is evident in the figure, and is a manifestation of the 
metallicity-luminosity relation in galaxies (e.g., Tremonti et al.
2004) and a strong metallicity dependence of the [\ion{N}{2}]/H$\alpha$ 
ratio (Pettini \& Pagel 2004).  We performed an ordinary 
least squares bisector fit to the 267 star-forming galaxies,
excluding galaxies with strong AGN contamination.
We imposed the additional constraint that the estimated value of 
[\ion{N}{2}]\,$\lambda6548,6584$/H$\alpha$ not exceed 0.54,
the approximate upper limit for HII regions and star-forming
galaxies in the BPT spectral sequence:\footnote{Two galaxies with measured integrated
[NII]/H$\alpha > 0.54$ (NGC~3351, NGC~4736) are listed with their measured values.}

\begin{equation}
\begin{array}{rl}
\mbox{log}(\mbox{[NII]}\lambda6548,6584/\mbox{H}\alpha) = (-0.173\pm0.007)\;M_B - (3.903\pm-0.137) &\mbox{if}\; M_B > -21\\
\mbox{[NII]}\lambda6548,6584/\mbox{H}\alpha =0.54 & \mbox{if}\; M_B \leq -21
\end{array}
\end{equation}

\noindent The 1-$\sigma$ scatter about this relation is $\pm$0.26 dex,
indicating that the estimates based on this correlation are only good to 
slightly better than a factor of two.  However since the 
[\ion{N}{2}]/H$\alpha$ ratios in the vast majority of galaxies
in our dwarf-dominated sample are very low ($<$0.1) the 
corresponding uncertainties in the total H$\alpha$ (or 
H$\alpha$ + [\ion{N}{2}]) fluxes is typically of order 10\%\
or less.
We have applied this scaling relationship to estimate the [NII]/H$\alpha$ ratio
whenever actual spectral measurements are not available from MK06, the 
Nearby Field Galaxies Survey (Jansen et al. 2000), 
or the various dwarf galaxy datasets of van Zee 
(van Zee \& Haynes 2006 and references therein).   


\pagebreak

\pagebreak
\begin{figure}[p]
\epsscale{1}
\plotone{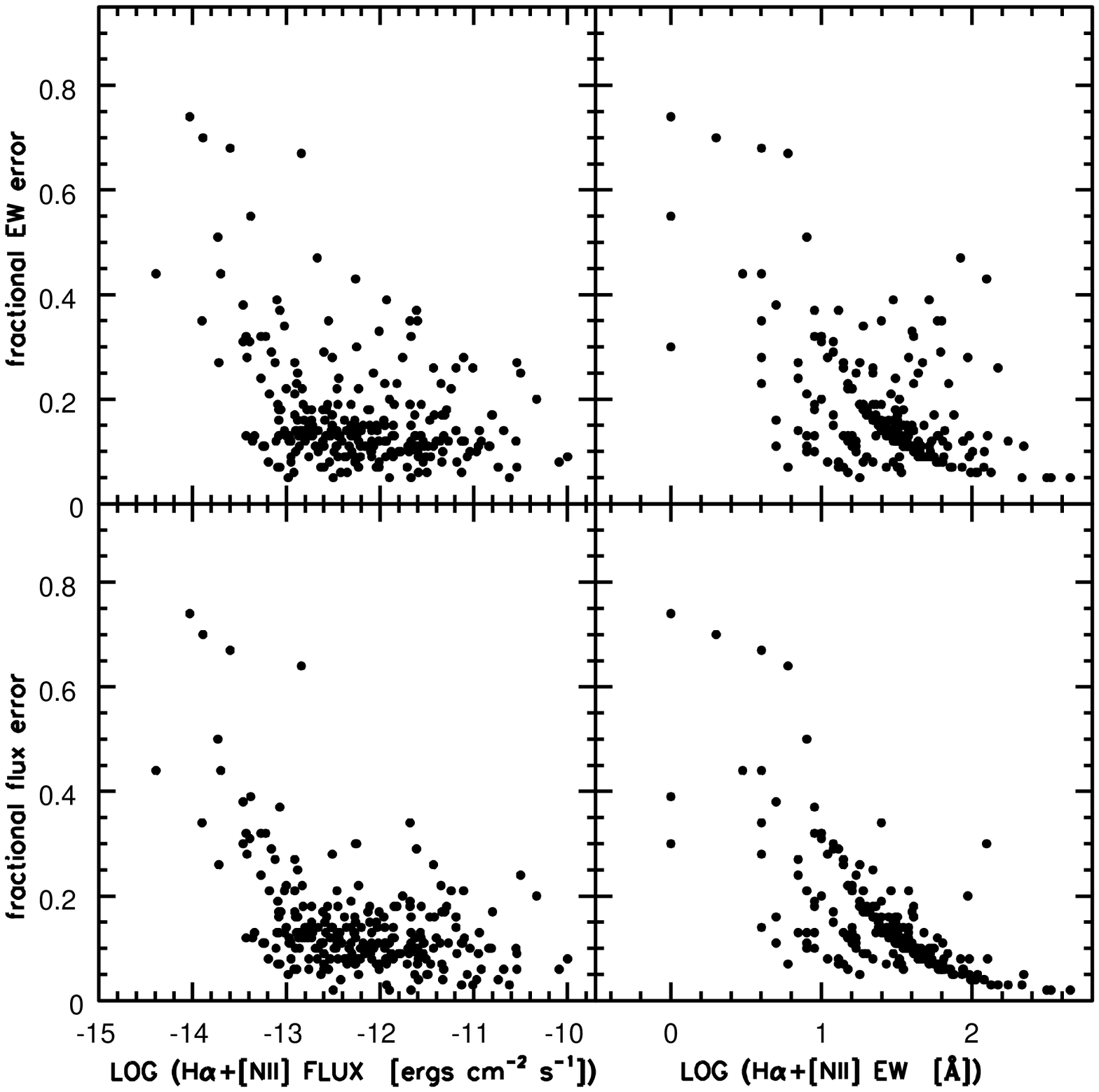}
\caption[Fractional Errors in the Integrated H$\alpha$ Fluxes 
and EWs]{Total flux and EW uncertainties plotted against 
the measurements themselves.  The errors generally increase 
with decreasing EW, due to the dominance of the continuum subtraction 
errors in the total uncertainty.  The fluxes and EWs have 
median uncertainties of approximately $\pm$12\%.}
\label{errors}
\end{figure}

\newpage
\begin{figure}[t]
\epsscale{1.1}
\plottwo{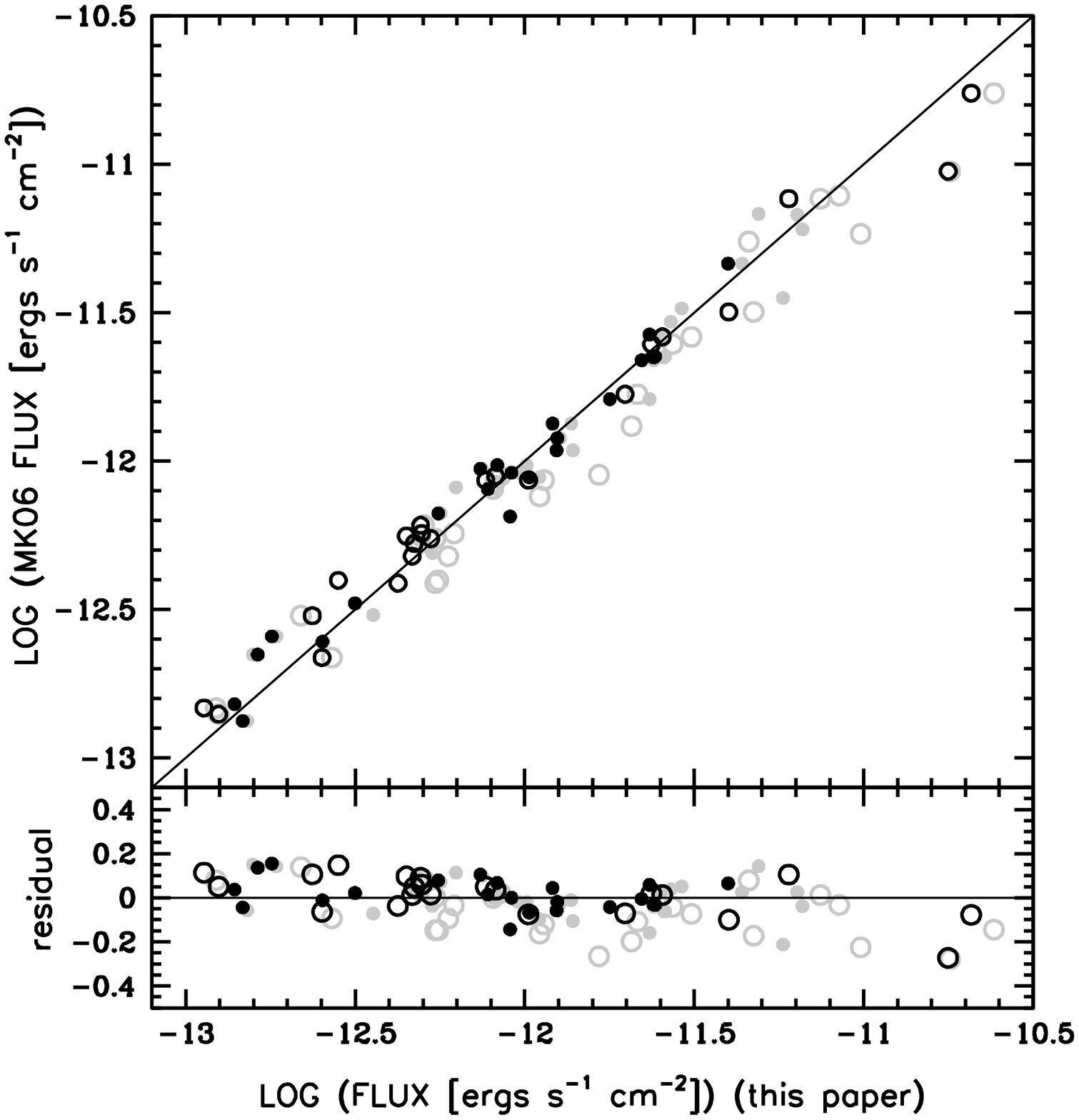}{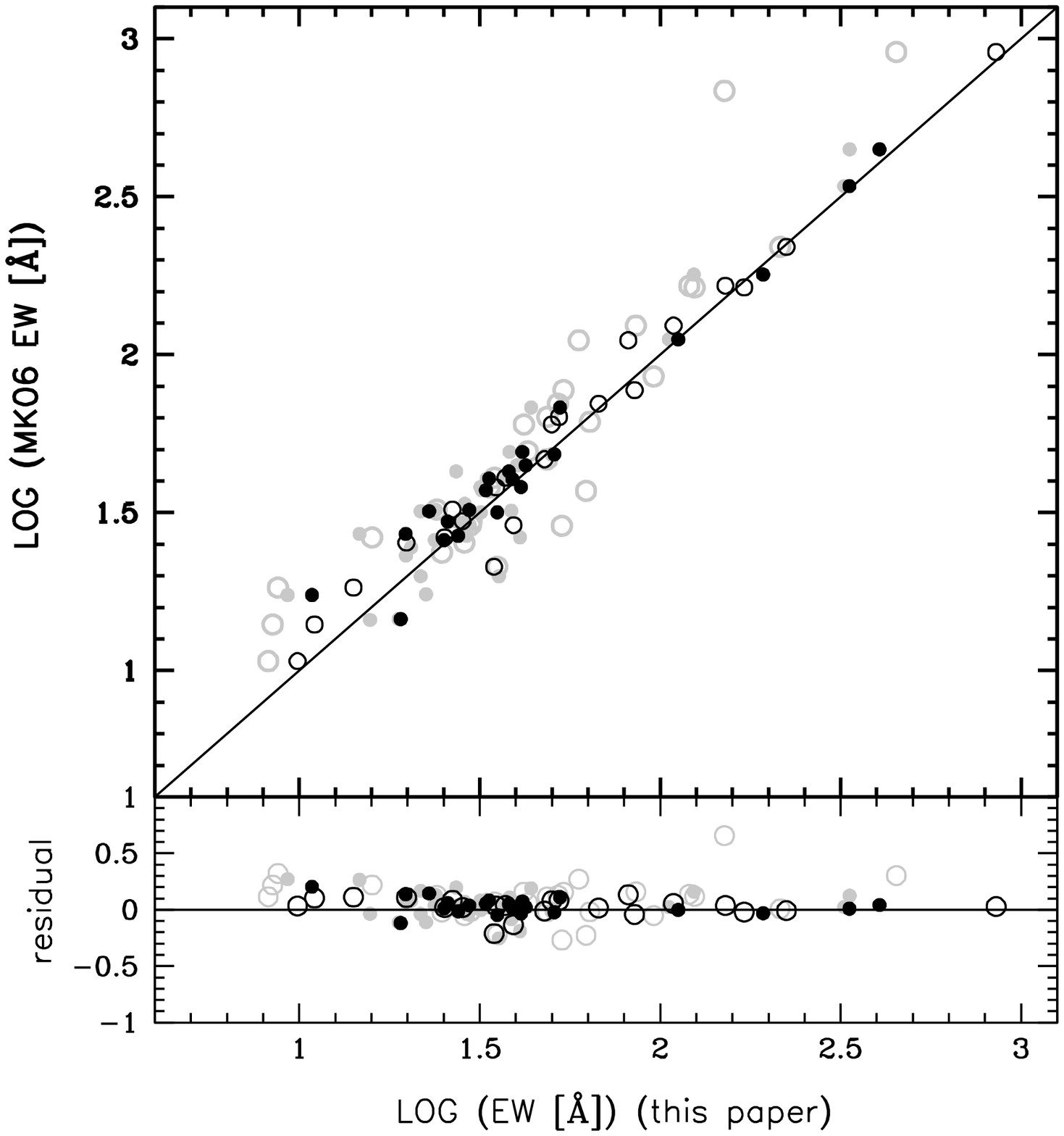}
\caption[Comparison to Integrated Spectral H$\alpha$ Flux and EW Measurements]
{({\it Left}) Integrated H$\alpha$+[NII] fluxes based on our 
observations plotted against measurements from the integrated spectral atlas 
of Moustakas \& Kennicutt (2006, MK06). Black symbols represent a comparison 
with imaging fluxes re-measured using the drift scan spectral apertures, 
while gray symbols represent a comparison that uses our original photometry.  
Open and filled circles denote galaxies where the extents of the nebular 
emission are limited by the MK06 apertures and those that are fully covered, 
respectively.  ({\it Right}) Same as the left panel, 
but for a comparison of the integrated H$\alpha$+[NII] EWs. }\label{kmcomp}
\end{figure}

\pagebreak
\begin{figure}[t]
\epsscale{1.1}
\plottwo{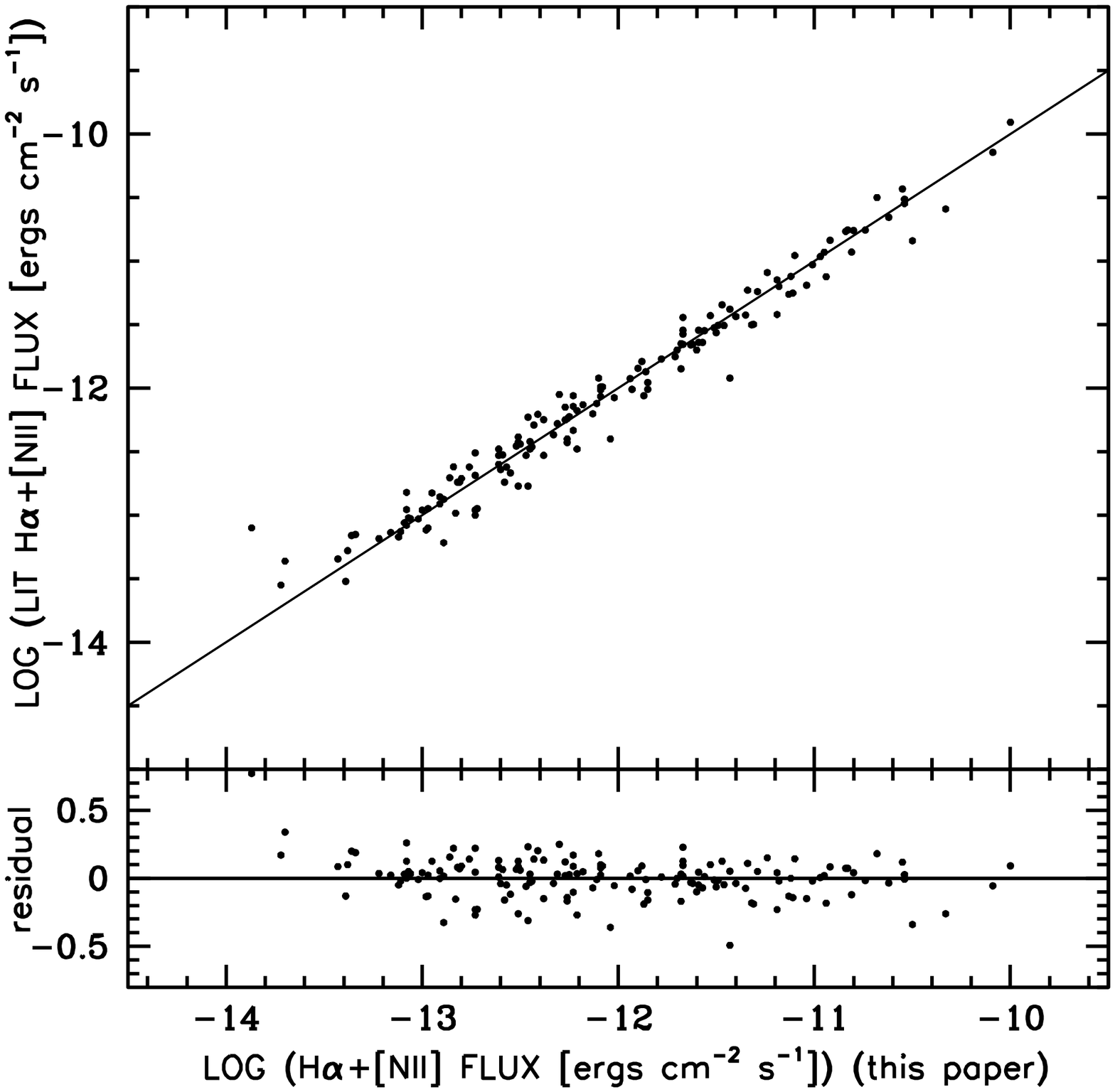}{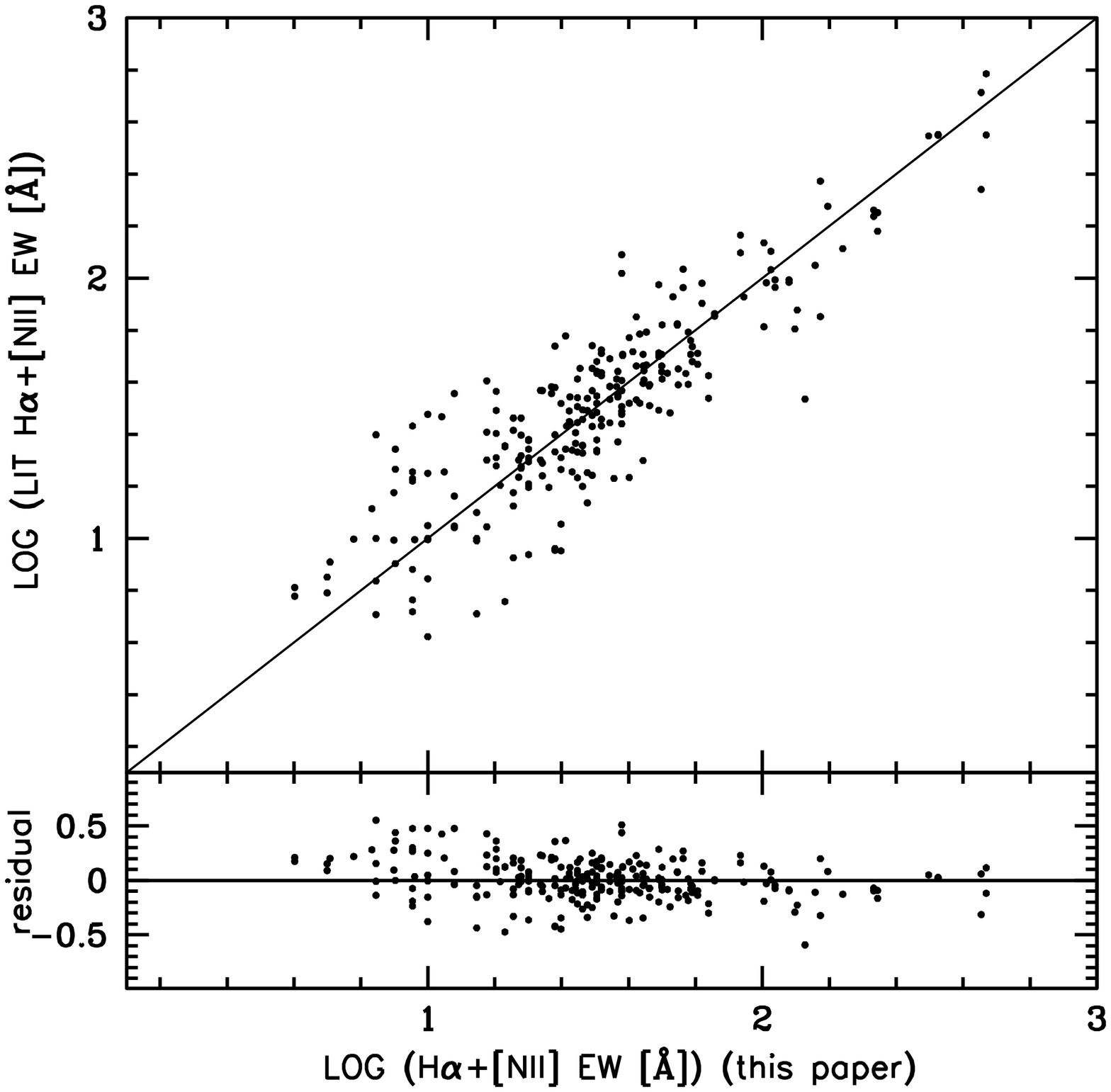}
\caption[Comparison to Additional H$\alpha$ Flux and EW Measurements in the 
Literature]{({\it Left}) Integrated H$\alpha$+[NII] fluxes 
from this paper compared to measurements from the homogenized flux 
compilation of Kennicutt et al. (2008, in preparation; KAL08).  
({\it Right})  Same as the previous panel, except that the
emission-line equivalent widths (EWs) are compared.}\label{litcomp}
\end{figure}

\pagebreak
\begin{figure}[t]
\epsscale{1}
\plotone{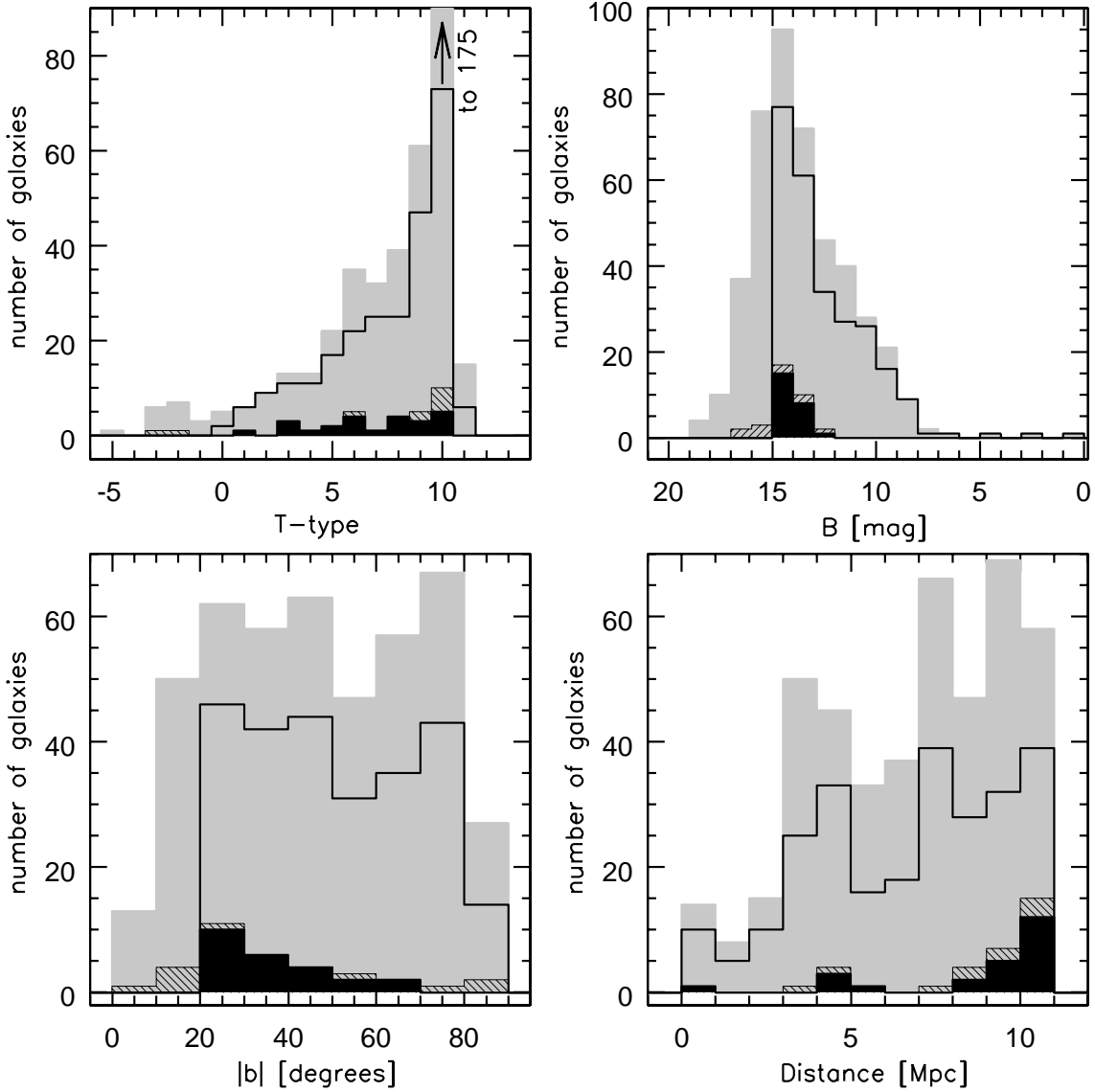}
\caption {Distributions of RC3 type (upper left), apparent $B$
magnitude (upper right), Galactic latitude (lower left), and 
distance (lower right) for the galaxies in this survey.  The shaded
distribution shows the properties for the entire galaxy sample
presented listed in Table 1.  The solid black lines show the
respective distributions for the primary sample, defined by
$T \ge 0$, $B \le 15$, $|b| \ge 20\degr$, and $d \le 11$ Mpc.
The small solid black and cross-hatched histograms denote
galaxies without H$\alpha$ photometry in the primary and
secondary subsamples, respectively.  For clarity the upper
left plot has been clipped at 90 galaxies; as indicated 
there are 167 galaxies with T = 10 (dwarf irregular) in the
entire sample.}
\end{figure}

\pagebreak
\begin{figure}[t]
\epsscale{1}
\plotone{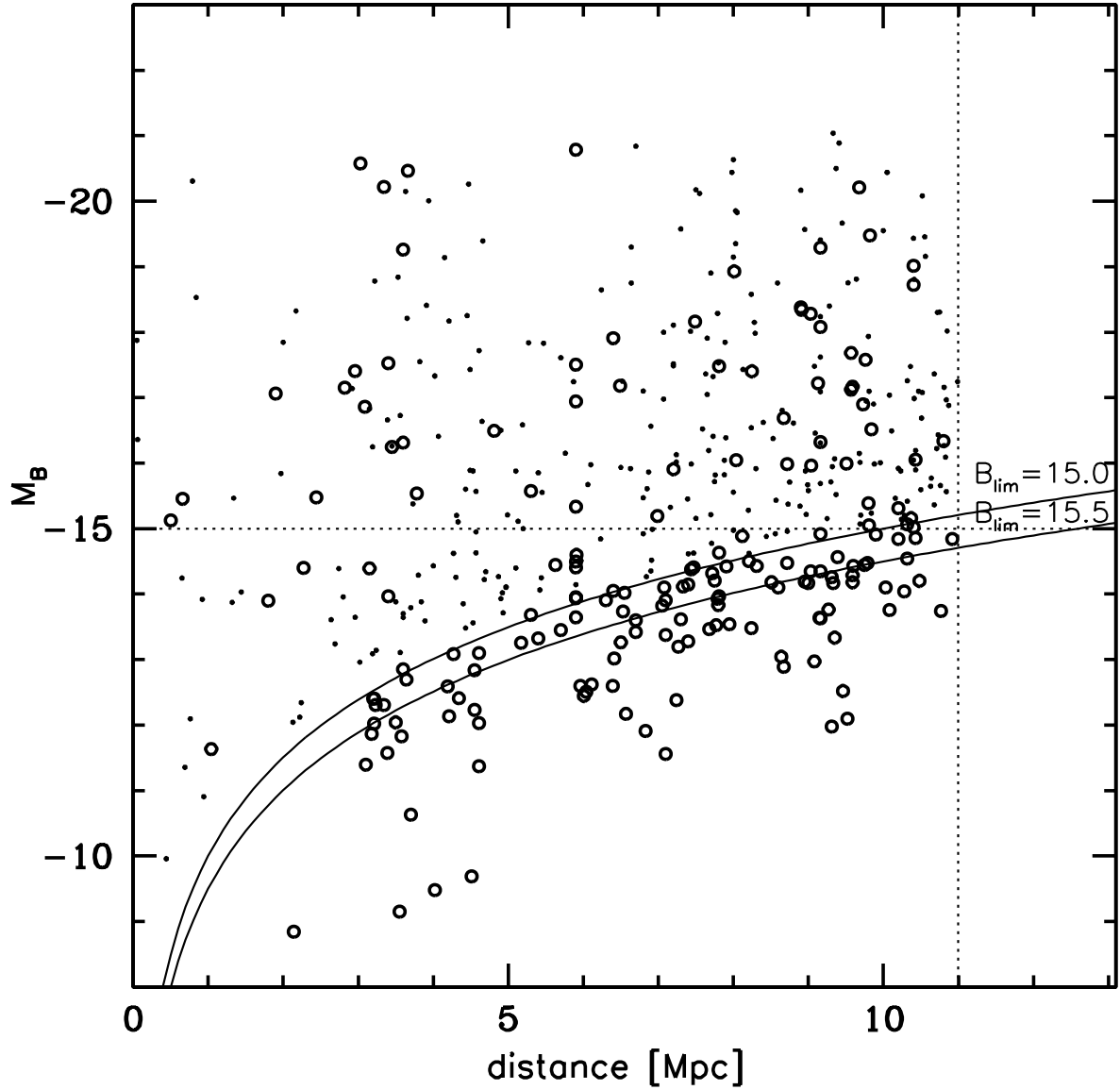}
\caption {Absolute blue magnitudes for the galaxies in this sample, plotted
as a function of distance.  Solid points denote galaxies in the
primary sample, while open circles show galaxies in the secondary
sample.  The two curves show lines of constant apparent magnitude
near the primary survey limit.}
\end{figure}

\pagebreak
\begin{figure}[t]
\epsscale{1}
\plotone{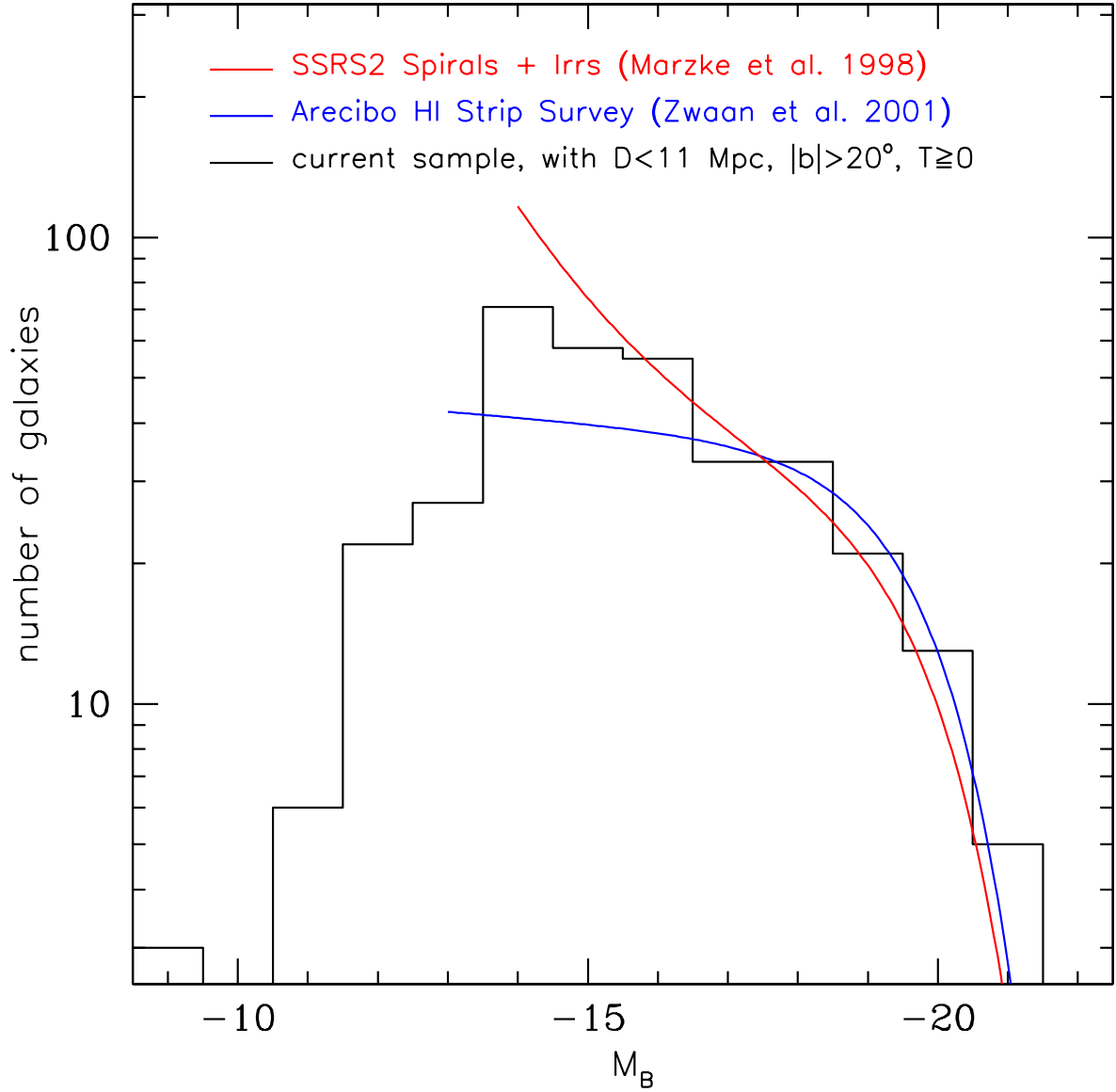}
\caption {Distribution of blue luminosities for the 11 Mpc 
sample, compared with published luminosity functions
taken from the optically-selected Second Southern Sky Redshift 
Survey (Marzke et al. 1998), and the HI-selected Arecibo Strip
Survey (Zwaan et al. 2001).  The luminosity functions are derived 
for spiral and irregular galaxies.}
\end{figure}

\pagebreak
\begin{figure}[t]
\includegraphics[scale=0.8]{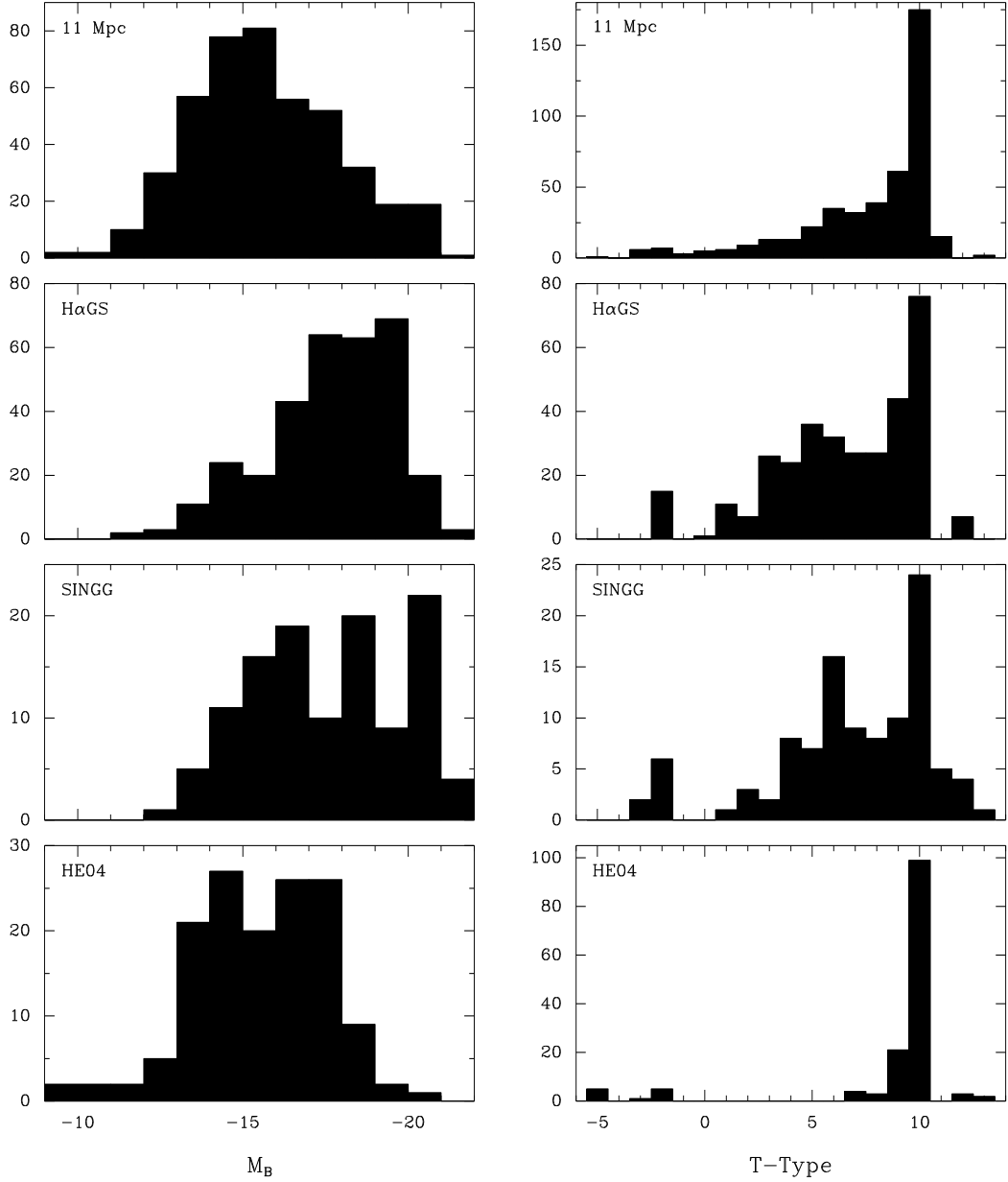}
\caption {Comparison of the distribution of Galactic extinction-corrected
absolute blue magnitudes
(left) and RC3 types for the galaxies in our sample (primary $+$ secondary
samples, top panels)
with 3 other large H$\alpha$ surveys, the H$\alpha$ Galaxy
Survey of James et al. (2004), the SINGG survey of Meurer et al. 
(2006), and the survey of dwarf irregular galaxies by Hunter \&
Elmegreen (2004), denoted in the bottom panel as HG04.}
\end{figure}

\pagebreak
\begin{figure}[t]
\includegraphics[scale=0.8]{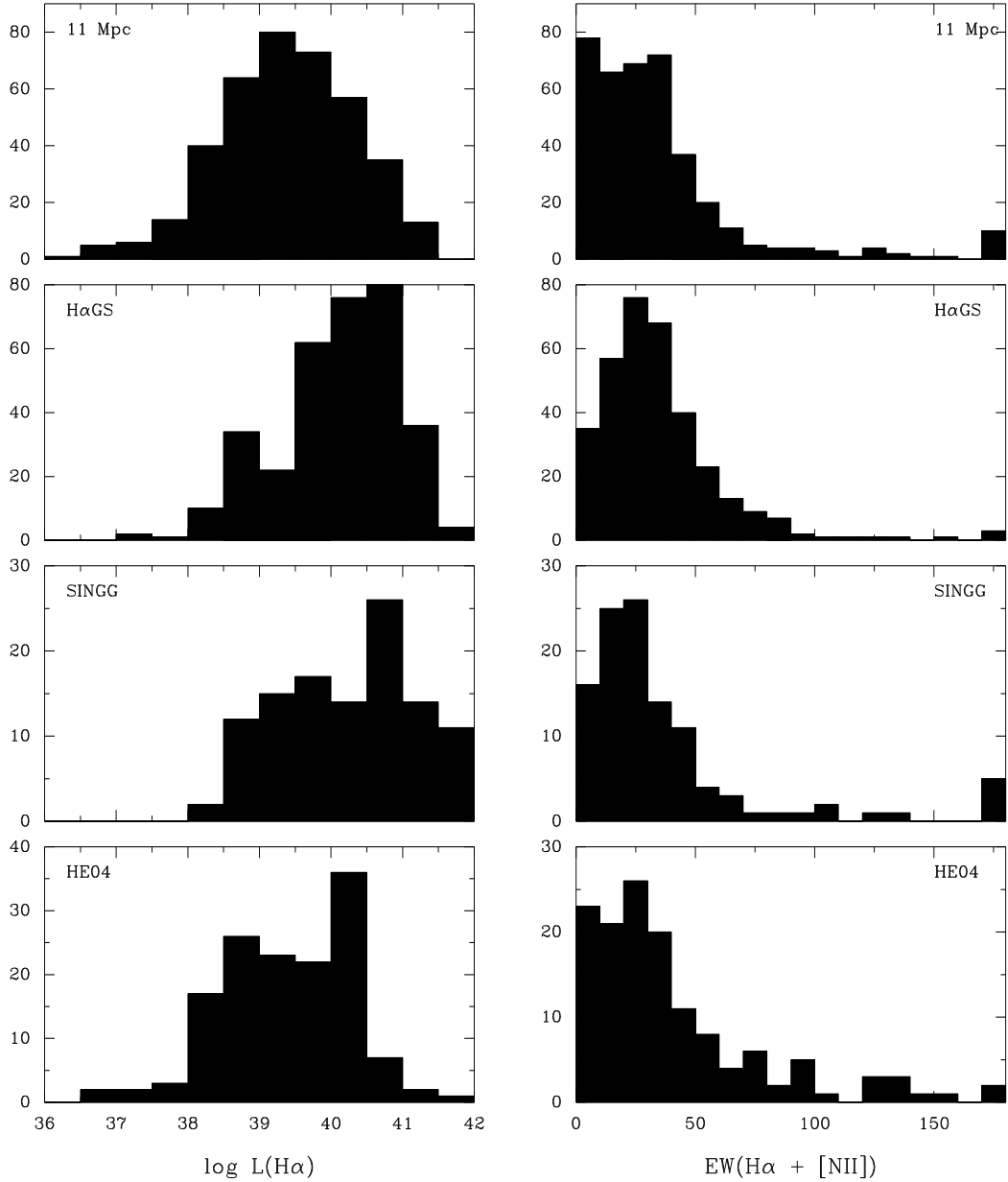}
\caption {Comparison of the distribution of H$\alpha$ + [\ion{N}{2}]
luminosities (in logarithmic units of ergs~s$^{-1}$) and EWs
(in \AA), for the same surveys as shown in Figure 7.  The
emission-line luminosities have been corrected for foreground
Galactic extinction, but no corrections for internal extinction
have been applied.  EWs for the HE04 galaxies were derived using
a combination of published H$\alpha$ fluxes and broadband magnitudes
(see text).  A handful of galaxies without detected line
emission have not been plotted, see text for a discussion of
these objects.}
\end{figure}

\pagebreak
\begin{figure}[t]
\epsscale{1}
\plotone{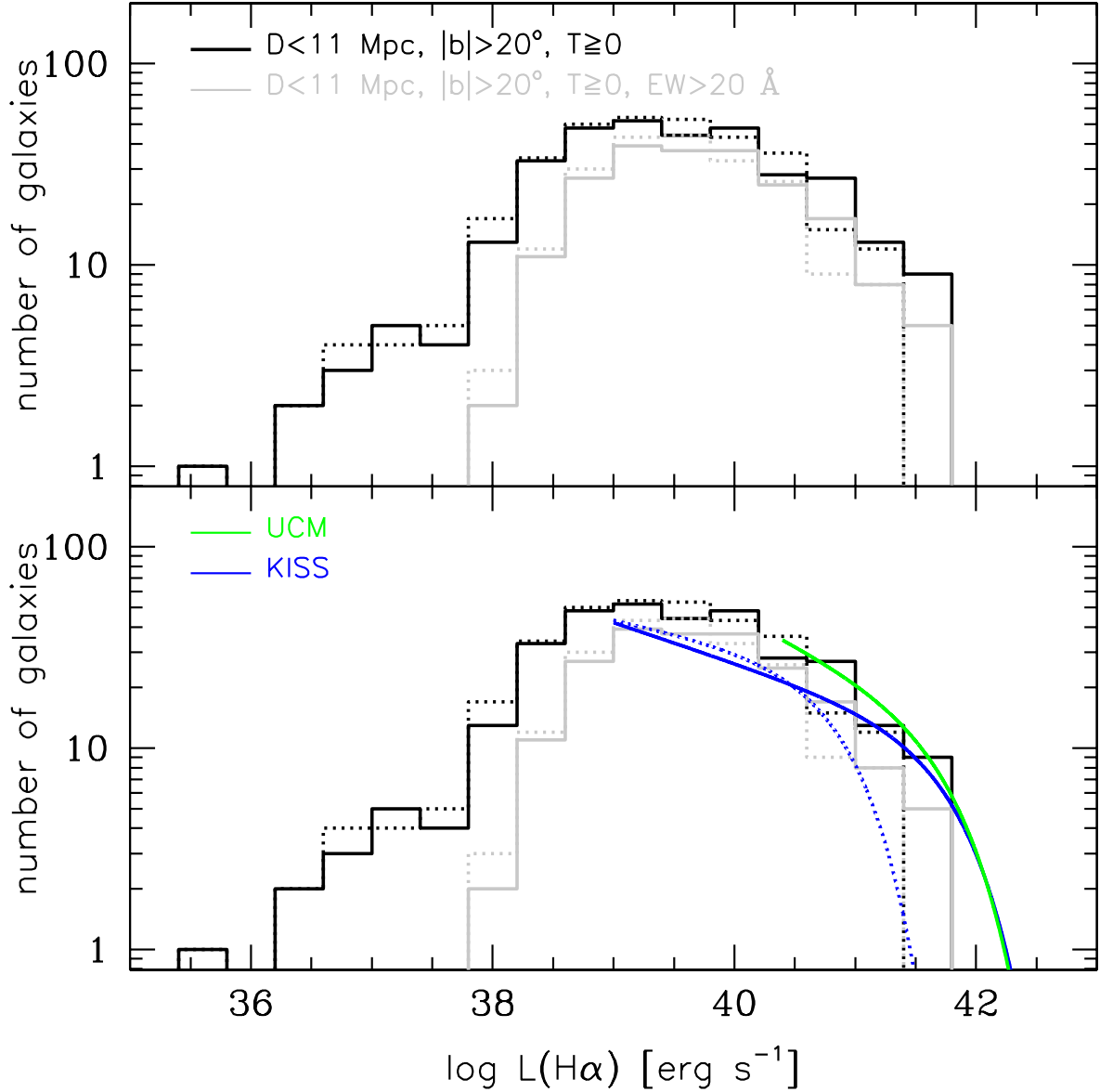}
\caption{{\it Upper panel}:  The H$\alpha$ emission-line luminosity function
for this sample.  Dashed histograms show the observed luminosities,
uncorrected for extinction, while the corresponding solid histograms 
show the extinction-corrected distributions.  The darker lines show
the luminosity function for the entire 11 Mpc sample for Galactic 
latitudes $|b| > 20$ and types S0/a and later, while the lighter
gray histograms show the same data, but with galaxies with 
EW(H$\alpha$+[\ion{N}{2}]) $<$ 20\,\AA\ excluded.  The latter
roughly replicates the typical EW completeness limit for published 
objective prism surveys (Gronwall et al. 2004).  
{\it Bottom panel}:  The same H$\alpha$ luminosity function
for our sample 
(with the EW cut applied) plotted as a histogram and compared to luminosity 
functions from the
UCM (green) and KISS (blue) wide-field objective prism surveys.
Dashed and solid lines show observed and extinction corrected
distributions, respectively.}
\end{figure}

\newpage
\begin{figure}[h]
\epsscale{0.8}
\plotone{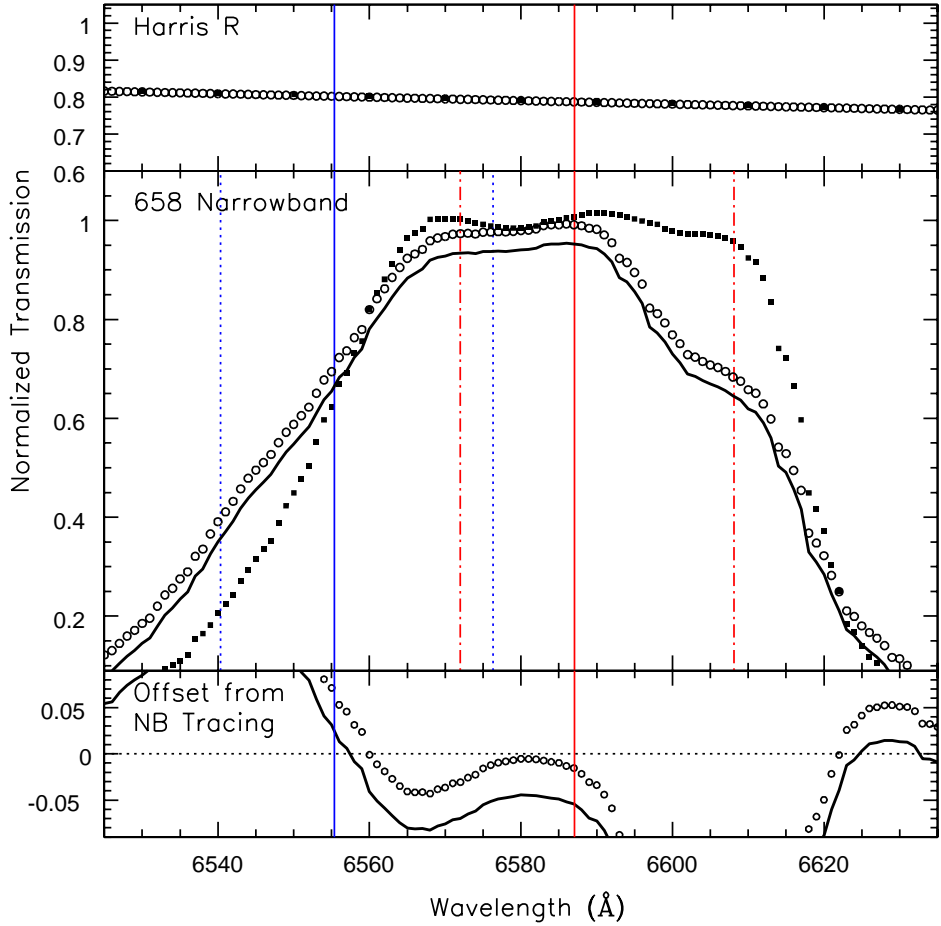}
\caption[Filter Response Functions and Transmission Corrections]
{Normalized response functions for the $R$-band 
filter used at the Bok telescope (top panel), and for the 
Andover 3-cavity interference ``658'' filter used for 
this survey (bottom panel).  The native transmission 
of the filters are traced by the solid squares.  The open 
circles represent the [NII] $+$ H$\alpha$ flux-weighted 
transmission, assuming the maximum [NII]$\lambda$6583/H$\alpha$ 
ratio of 0.54 for star-forming galaxies (see Appendix B).  
The solid curve describes the final effective 658 
transmission which has further been corrected for
the loss of emission-line flux during continuum subtraction.  
The extents of these corrections are illustrated with the 
residual plot in the bottom panel.  The solid red and blue 
lines mark the wavelength range over which H$\alpha$ is shifted 
for the recessional velocities spanned by the galaxies in our 
sample.  The corresponding positions of the [NII]$\lambda\lambda$6548,6583 
lines are also indicated in the middle panel. }\label{eff_trans}
\end{figure}

\pagebreak
\begin{figure}[h]
\epsscale{0.8}
\plotone{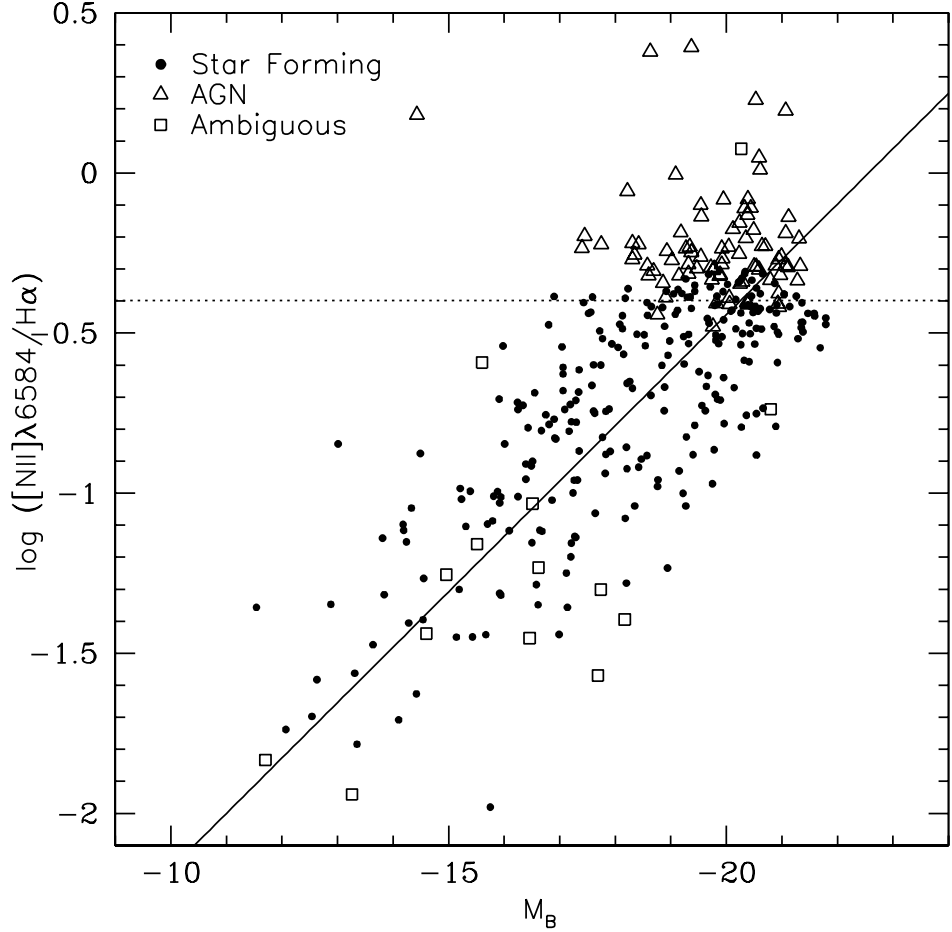}
\caption[NII $\lambda$6584/H$\alpha$ vs. $M_B$ from Integrated Spectral 
Measurements]{The [NII]$\lambda$6584/H$\alpha$ 
ratio plotted against $M_B$ for galaxies in the MK06 
integrated spectral atlas (note that the total doublet strength 
[NII]$\lambda$6548,6584/H$\alpha$ is higher by a factor 1.33).  Points represent star-forming 
galaxies while triangles denote galaxies with AGN.  Squares represent 
galaxies with ambiguous classifications.  The solid line shows 
the best fit line, excluding those galaxies with AGN.  The fit is 
used to correct the observed fluxes for [NII] contamination when 
individual estimates of the [NII]/H$\alpha$ ratio from the 
integrated spectra of MK06 and the NFGS, or the various spectral datasets 
of van Zee et al. (van Zee \& Haynes 2006 and references therein) are not 
available.}\label{niiha}
\end{figure}

\newpage


\begin{deluxetable}{clcrrrrllccccc}
\tabletypesize{\scriptsize}
\rotate
\tablecolumns{14}
\tablewidth{0pc}
\tablecaption{Galaxy Sample}

\tablehead{
\colhead{\#}            &
\colhead{Galaxy Name}   &
\colhead{Sample}        &
\colhead{RA}            &
\colhead{DEC}           &
\colhead{$b$}           &
\colhead{T}             &
\colhead{B}             &
\colhead{err(B)}        &
\colhead{$cz$}          &
\colhead{v(LG) }      &
\colhead{D}             &
\colhead{method}        &
\colhead{References}    \\

\colhead{}              &  
\colhead{}              &
\colhead{Flag}          &
\colhead{[J2000]}       &
\colhead{[J2000]}       &
\colhead{}              &
\colhead{}              &
\colhead{[mag]}         &
\colhead{[mag]}         &
\colhead{[km s$^{-1}$]} &
\colhead{[km s$^{-1}$]} &
\colhead{[Mpc]}         &
\colhead{}              &
\colhead{}    \\

\colhead{(1) }           &  
\colhead{(2) }           &
\colhead{(3) }           &
\colhead{(4) }           &
\colhead{(5) }           &
\colhead{(6) }           &
\colhead{(7) }           &
\colhead{(8) }           &
\colhead{(9) }           &
\colhead{(10) }          &
\colhead{(11) }          &
\colhead{(12) }          &
\colhead{(13) }          &
\colhead{(14) }          
}
\startdata
    1 & UGC12894 & s &000022.5 & 392944 & -22.32 & 10 & 16.55 &  0.20 & 335 &  618 &  8.2 & v(flow) & 120\\ 
     2 & WLM & p &000158.1 & -152739 & -73.63 & 10 & 11.03 &  0.08 &-122 &  -16 & 0.92 & trgb & 11, 118\\ 
     3 & ESO409-IG015 & s &000531.8 & -280553 & -79.79 &  6 & 15.14 &  0.19 & 737 &  780 & 10.4 & v(flow) & 111\\ 
     4 & ESO349-G031 & s &000813.3 & -343442 & -78.12 & 10 & 15.56 &  0.20 & 221 &  230 & 3.21 & trgb &  7, 111\\ 
     5 & NGC24 & p &000956.7 & -245744 & -80.43 &  5 & 12.19 &  0.13 & 554 &  610 &  8.1 & v(flow) & 118\\ 
     6 & NGC45 & p &001404.0 & -231055 & -80.67 &  8 & 11.32 &  0.08 & 467 &  530 &  7.1 & v(flow) & 118\\ 
     7 & NGC55 & p &001454.0 & -391149 & -75.74 &  9 &  8.42 &  0.05 & 129 &  111 & 2.17 & trgb & 62, 118\\ 
     8 & NGC59 & s &001525.4 & -212642 & -80.02 & -3 & 13.12 &  0.14 & 362 &  432 &  5.3 & sbf & 21, 118\\
\enddata
\end{deluxetable}


\newpage

\begin{deluxetable}{llccccr}
\tabletypesize{\footnotesize}
\tablecaption{CCD Imager Properties}
\tablewidth{0pt}
\tablehead{
\colhead{Telescope} & 
\colhead{Detector}  & 
\colhead{CCD Scale} & 
\colhead{FOV} & 
\colhead{Continuum Filter} & 
\colhead{Line Filter(s)} &
\colhead{Exp Times}\\
\colhead{(1)}&
\colhead{(2)}&
\colhead{(3)}&
\colhead{(4)}&
\colhead{(5)}&
\colhead{(6)}& 
\colhead{(7)} \\
}
\startdata
Bok 2.3~m & Loral 2K$\times$2K  & 0\farcs43      & 4\farcm9      &6451/1473     &6585/66 & 1000/200 \\

VATT 1.8m & Loral 2K$\times$2K  & 0\farcs40      & 6\farcm4      &6338/1186     &6585/66 & 1800/360 \\
          &                     &           &               &              &6600/69 \\

CTIO 0.9m & Tek 2K$\times$2K    & 0\farcs79      & 13\farcm5     &6425/1500     &6563/75 & 2700/360 \\
          &                     &           &               &              &6600/75 \\
\enddata
\tablecomments{Col(1): Telescope and abbreviation used in the paper;
  Col(2): CCD detector;  
  Col(3): CCD scale in arcsec~pixel$^{-1}$;
  Col(4): Detector field of view (arcminutes);
  Col(5): Continuum band ($R$) central wavelength and FWHM bandwidth, in \AA;
  Col(6): H$\alpha$+[NII] filter central wavelength and FWHM bandpass (\AA);
  Col(7): Standard exposure times narrowband/continuum filters (seconds)}
\end{deluxetable}

\newpage


\begin{deluxetable}{clr@{}c@{}rr@{}r@{}rrrccccccc}
\tabletypesize{\scriptsize}
\rotate
\tablecolumns{16}
\tablewidth{0pc}
\tablecaption{Integrated H$\alpha$+[NII] Fluxes and Equivalent Widths}

\tablehead{
\colhead{\#}                 &
\colhead{Galaxy Name}        &
\multicolumn{3}{c}{log $f$}  &
\multicolumn{3}{c}{EW}  &
\colhead{Source}             &
\colhead{Telescope}          &
\colhead{Coverage}           &
\colhead{$A_b$}              &
\colhead{$M_B$}              &
\colhead{[NII]/H$\alpha$}    &
\colhead{Ref.}               &
\colhead{L(H$\alpha$)}       \\

\colhead{(1)}           &  
\colhead{(2)}           &
\multicolumn{3}{c}{(3)}   &
\multicolumn{3}{c}{(4)}   &
\colhead{(5)}           &
\colhead{(6)}           &
\colhead{(7)}           &
\colhead{(8)}           &
\colhead{(9)}           &
\colhead{(10)}          &
\colhead{(11)}          &
\colhead{(12)}

}
\startdata
    1 & UGC12894 &-13.42 & $\pm$ & 0.03 & \nodata & \nodata & \nodata & 
 2.3 & \nodata & \nodata &   0.45 & 
 -13.48 &   0.04 & 3 & 38.58\\ 
     2 & WLM &-11.68 & $\pm$ & 0.13 & 25 & $\pm$ & 9 & 
 1.1 & KPNO & 2.2 &   0.13 & 
 -13.91 &   0.04 & 3 & 38.34\\ 
     3 & ESO409-IG015 &-12.34 & $\pm$ & 0.04 & 262 & $\pm$ & 26 & 
 2.1 & \nodata & \nodata &   0.08 & 
 -15.02 &   0.07 & 3 & 39.77\\ 
     4 & ESO349-G031 &$<$-15.10 & \nodata & \nodata & 0 & \nodata & \nodata & 
 2.2 & \nodata & \nodata &   0.05 & 
 -12.03 &   0.02 & 3 & $<$36.00\\ 
     5 & NGC24 &-11.87 & $\pm$ & 0.04 & 19 & $\pm$ & 1 & 
 2.1 & \nodata & \nodata &   0.07 & 
 -17.43 &   0.17 & 3 & 39.98\\ 
     6 & NGC45 &-11.22 & $\pm$ & 0.04 & 38 & $\pm$ & 10 & 
 2.2 & \nodata & \nodata &   0.07 & 
 -18.00 &   0.22 & 3 & 40.49\\ 
     7 & NGC55 &-10.12 & $\pm$ & 0.02 & 41 & $\pm$ & 10 & 
 2.2 & \nodata & \nodata &   0.06 & 
 -18.32 &   0.25 & 3 & 40.55\\ 
     8 & NGC59 &-12.36 & $\pm$ & 0.06 & 38 & $\pm$ & 5 & 
 1.2 & KPNO & 1 &   0.07 & 
 -15.58 &   0.08 & 3 & 39.15\\
\enddata
\tablenotetext{1}{The fluxes for NGC 1800 and its companion MCG-05-13-004 were
measured together within a single aperture.}
\end{deluxetable}


\end{document}